\documentclass[12pt]{article}
\usepackage{graphicx}
\usepackage{latexsym}
\advance\oddsidemargin by -\textwidth
\advance\oddsidemargin by -1in
\evensidemargin=\oddsidemargin
\def\fsl#1{\setbox0=\hbox{$#1$}           
   \dimen0=\wd0                                 
   \setbox1=\hbox{/} \dimen1=\wd1               
   \ifdim\dimen0>\dimen1                        
      \rlap{\hbox to \dimen0{\hfil/\hfil}}      
      #1                                        
   \else                                        
      \rlap{\hbox to \dimen1{\hfil$#1$\hfil}}   
      /                                         
   \fi}                                         %
\makeatletter
\def\@maketitle{\newpage
 \null
 {\normalsize \tt \begin{flushright} 
  \begin{tabular}[t]{l} \@date 
  \end{tabular}
 \end{flushright}}
 \begin{center}
 \vskip 2em
 {\LARGE \@title \par} \vskip 2.5em {\large \lineskip .5em
 \begin{tabular}[t]{c}\@author 
 \end{tabular}\par} 
 \end{center}
} 

\newcommand{\vereq}[2]{\lower3pt\vbox{\baselineskip1.5pt \lineskip1.5pt
\ialign{$\m@th#1\hfill##\hfil$\crcr#2\crcr\sim\crcr}}}
\newcommand{\lesssim}{\mathrel{\mathpalette\vereq<}}

\newcommand\eqsecnum{
\@newctr{equation}[section]
\renewcommand\theequation{\arabic{section}.\arabic{equation}}%
}

\newbox\tempboxa
\newdimen\captionboxsubcount 
\def\capsize#1{\captionboxsubcount=#1pt}
\newdimen\captionboxsub
\captionboxsub=\hsize \advance\captionboxsub by -\captionboxsubcount
\advance\captionboxsub by -\captionboxsubcount
\long\def\@makecaption#1#2{
 \setbox\@tempboxa\hbox{\footnotesize #1: #2}
 \ifdim \wd\@tempboxa >\captionboxsub 
\rightskip=\captionboxsubcount \leftskip=\captionboxsubcount 
  \footnotesize #1: #2 
\else \hbox to\hsize{\hfil\box\@tempboxa\hfil} 
 \fi}
\makeatother
\eqsecnum
\capsize{30}
\title{{\Large\bf 
   Meson masses in large $N_f$ QCD \\
     from the Bethe-Salpeter equation\vspace{5mm}}}
\author{
{\large
    Masayasu {\sc Harada}\thanks{
      {\tt harada@eken.phys.nagoya-u.ac.jp}}, \ \ 
    Masafumi {\sc Kurachi}\thanks{
      {\tt kurachi@eken.phys.nagoya-u.ac.jp}} \ \  and \ \ 
    Koichi {\sc Yamawaki}\thanks{
      {\tt yamawaki@eken.phys.nagoya-u.ac.jp}}
  }\\[3mm]
  {\it Department of Physics, Nagoya University}\\
  {\it Nagoya 464-8602, Japan}\\[1cm]
}
\date{
  DPNU-03-09 \vspace{1cm}\\
}
\begin{document}
\maketitle

\vspace{1cm}
\begin{abstract}
We solve the homogeneous Bethe-Salpeter (HBS) equation 
for the scalar, pseudoscalar, vector, and axial-vector
bound states of quark and anti-quark 
in large $N_f$ QCD with the improved ladder 
approximation in the Landau gauge. 
The quark mass function in the HBS equation is obtained from the 
Schwinger-Dyson (SD) equation in the same approximation for  
consistency with the chiral symmetry.
Amazingly, due to the fact that the two-loop running coupling 
of large $N_f$ QCD is explicitly written in terms of an analytic function,
large $N_f$ QCD turns out to be the first example in which 
the SD equation can be solved 
in the complex plane and hence
the HBS equation directly in the time-like region.
We find that approaching the chiral phase transition point from the broken
phase,  the scalar, vector, and axial-vector meson masses 
vanish to zero  with the same scaling behavior, all degenerate with the massless
pseudoscalar meson.
This may suggest a new type of manifestation of the chiral symmetry
restoration in large $N_f$ QCD. 
\end{abstract}

\newpage
\section{Introduction}
\label{sec:Introduction}

Spontaneous chiral symmetry breaking is one of the
most important properties to understand the low-energy phenomena
of QCD in the real world.
This chiral symmetry is expected to be restored in QCD at several 
extreme conditions such as QCD with a large number of massless quarks,
large $N_f$ QCD (see, e.g., Refs.~\cite{Banks:nn,Appelquist:1996dq,
Miransky:1996pd,Appelquist:1998rb,lattice,OZ,VS}) and
QCD in hot and/or dense matter (see, e.g., Ref.~\cite{hot-dense}).
In Ref.~\cite{Appelquist:1996dq},
based on the infrared (IR) fixed point existing at
a two-loop beta function for a large number of massless quarks
($N_f \lesssim\frac{11}{2}N_c$)~\cite{Banks:nn}, it was found
through the 
improved 
ladder Schwinger-Dyson (SD) equation
that chiral symmetry restoration 
takes place for $N_f$ 
such that $N_f^{\rm crit} < N_f < \frac{11}{2} N_c$, where
$N_f^{\rm crit}\simeq 4 N_c$($=12$ for $N_c=3$).
Then, in Ref.~\cite{Miransky:1996pd}
this chiral restoration at $N_f^{\rm crit}$
was further identified with 
``conformal phase transition'' which was characterized by the 
essential singularity scaling.
Moreover, such chiral restoration is also observed 
by other various methods such as lattice
simulation~\cite{lattice}, dispersion relation~\cite{OZ},
instanton calculus~\cite{VS}, 
effective field theoretical approach~\cite{Harada:1999zj},
etc.. 

More attention has been paid 
to the property of the phase transition.
Especially, it is interesting to ask what are the
light degrees of 
freedom
near the phase transition point in the
large $N_f$ QCD:
For example, in the manifestation of the chiral symmetry restoration 
\'{a} la the linear sigma model, 
the scalar 
bound state becomes a chiral partner of the pseudoscalar bound state 
and becomes massless at the phase transition point.
On the contrary, in the vector manifestation
(VM)~\cite{Harada:2000kb,HY:PRep} obtained by the
effective field theoretical approach based on the hidden local
symmetry model~\cite{Bando:1984ej}, 
it is the vector
bound state which becomes massless as a chiral partner of the
pseudoscalar bound state.
Besides, from the viewpoint of the conformal phase 
transition~\cite{Miransky:1996pd}, 
it is natural to suppose that 
all the existing bound states become massless near the phase
transition point when approached from the broken 
phase (see Ref.~\cite{Chivukula:1996kg}).

Then, it is quite interesting to study which types of the bound states
actually exist near the phase transition point, and 
investigate the critical behavior of their masses directly from QCD.
Such studies from the first principle
will give us a clue to understand the nature of the chiral phase
transition in large $N_f$ QCD.

A powerful tool to study the bound states of quark and anti-quark
directly from QCD is the
homogeneous Bethe-Salpeter (HBS) equation
in the (improved) ladder approximation
(see, for reviews,
Refs.~\cite{Nakanishi:ph,Kugo:review,Miransky:book}).
When the mass of the quark is regarded as a constant,
we can easily solve the HBS equation by using a so-called
fictitious eigenvalue method~\cite{Nakanishi:ph}.
However, 
for consistency with the chiral symmetry,
the quark propagator in the HBS equation
must be obtained by solving the 
SD equation 
with the same kernel as that used in the HBS 
equation~\cite{MN,Kugo:1992pr,Kugo:1992zg,Bando-Harada-Kugo}, 
and as a result,
the quark mass becomes a certain momentum dependent
function.
Then, in order to obtain the masses and the wave functions 
of the bound states, it is necessary to solve the HBS equation 
and the SD equation simultaneously.

When we try to solve these two
equations in 
real-life ($N_f = 3$) QCD,
however, we encounter difficulties.
First of all, for the consistency of the solution of the SD
equation with QCD in a high energy region,
we need to use the running coupling which obeys the evolution 
determined from QCD $\beta$-function in the high energy 
region (see, for reviews, Refs.~\cite{Kugo:review,Miransky:book}).
Since the 
running coupling diverges at some infrared scale, $\Lambda_{\rm QCD}$,
we have to regularize the running coupling in
the low energy region, 
for which there exist several ways
(see, e.g., Refs.~\cite{Higashijima:1983gx,Miransky:vj,
Aoki:1990eq,Aoki:1990aq,Roberts-Schmidt}).
Even if we fix the infrared regularization in such a way
that we can solve the 
SD equation on the real (space-like) axis,
another problem arises when we try to solve the HBS equation.
Since the argument of the quark mass function in the HBS equations for
the massive bound states becomes a complex quantity after the Wick 
rotation has been made, we have to solve the SD equation on the
complex plane, which requires an analytic continuation of the 
running coupling.
Several works such as
in Refs.~\cite{Kugo-Mitchard-Yoshida,Maris-Roberts-Tandy}
proposed models of running couplings for a general complex variable
which are consistent
with perturbative QCD for large space-like momentum.
However, they still have branch cuts on the complex plane,
and it is a complicated task to obtain the solution of the SD equation
for a general complex variable.
One way to avoide a such complication 
is solving the inhomogeneous BS equation for vertex functions
to obtain the current correlators in the space-like region
which we can fit the mass of the relevant bound state to
(see, e.g., Refs.~\cite{Aoki:1990yp,Naito-Oka}).
Another way 
might be replacing 
the entire running coupling with  
an ad hoc analytic function 
(see, e.g., Ref.~\cite{Alkofer-Watson-Weigel}).
Anyway, it is impossible to solve the SD equation on the complex plane 
without modeling the running coupling. 

In this paper, we point out that the situation dramatically changes 
when we increase the number of massless quarks.
When $N_f$ becomes larger than $\ N_f^\ast \simeq 8.05$,
the running coupling obtained from the renormalization group equation 
(RGE) with two-loop approximation takes a finite value for all the
range of the energy region due to the emergence of the non-trivial
IR fixed point. 
Then, we need no infrared regularization, and we do not have any 
ambiguities coming from the regularization scheme 
which do 
exist
in the case of small $N_f$.
Moreover, an explicit solution of the two-loop RGE can be written 
in terms of the Lambert $W$ 
function~\cite{Appelquist:1998rb,Gardi}, and 
when $N_f$ is close to $N_f^{\rm crit}$ 
the solution of the RGE has 
no singularity on the complex plane except for the time-like 
axis~\cite{Gardi}.
Consequently, we can solve the SD equation on the complex plane 
without introducing any models for the running coupling.

Based on these facts, 
we solve the HBS equations for the bound states 
of quark and anti-quark in large $N_f$ QCD 
with the improved ladder approximation in the 
Landau gauge.
The mass function for complex arguments needed
in the HBS equation is obtained by
solving the SD equation with the same kernel as that used in the HBS
equation.
We find the solution of the HBS equation in each of
the scalar, vector, and axial-vector channels,
which implies that 
the scalar, vector, and axial-vector bound states are 
actually formed near the 
phase transition point.
Our results show that the masses of the scalar, 
vector, and axial-vector  bound states go to zero as the number of 
quarks $N_f$ approaches to its critical value $N_f^{\rm crit}$ 
where the chiral symmetry restoration takes place.
This may suggest the existence of a new type of manifestation of
chiral symmetry restoration in large $N_f$ QCD
other than the linear sigma model like
manifestation and a simple version of the
vector manifestation proposed in Ref.~\cite{Harada:2000kb}.

This paper is organized as follows.
In section~\ref{sec:SDeq} we numerically solve the SD equation
with an approximate form of the running coupling,
and study the critical behavior of 
the Nambu-Goldstone boson decay constant.
In section~\ref{sec:complex_SD}
we solve the SD equation 
for complex arguments.
Section~\ref{sec:HBSeq} is devoted to
summarizing the numerical method for solving the HBS equation.
Section~\ref{sec:Numerical_calculations} is the main part of this
paper.
We first solve the HBS equation for the pseudoscalar bound state
to show
that the approximation adopted in the present analysis is 
consistent with the chiral symmetry.
We next solve the HBS equation for 
the scalar, vector, and 
axial-vector bound states to obtain
their masses.
Finally we give a summary and discussion in 
section~\ref{sec:Discussion_and_conclusion}.
In Appendix~\ref{app:positronium} we solve the HBS equation for the
ortho-positronium with a constant electron mass to show the validity
of the fictitious eigenvalue method.
The bispinor bases for the bound states are listed in 
Appendix~\ref{app:bispinor-bases}. 
In Appendix~\ref{app:decay_const} we calculate the coupling constants 
$F_V$, $F_A$, and $G_S$ of the vector, axial-vector, and scalar
bound states to the vector current, axial-vector current, and scalar
density.
We briefly study numerical uncertainties in the present analysis
in Appendix~\ref{app:Uncertainties}.

\section{Schwinger-Dyson equation in large $N_f$ QCD}
\label{sec:SDeq}

In this section we 
numerically solve
the Schwinger-Dyson (SD) equation 
for the quark propagator
with the improved ladder
approximation in the Landau gauge,
and show the critical behaviors of the dynamical 
mass and the decay constant of the Nambu-Goldstone boson.
We also show the behavior of the fermion-antifermion pair 
condensate $\ \langle \bar{\psi} \psi \rangle\ $ near the phase 
transition point.

\subsection{SD equation in the (improved) ladder approximation}
\label{sec:ladderSD}
\begin{figure}
  \begin{center}
    \includegraphics[height=3.3cm]{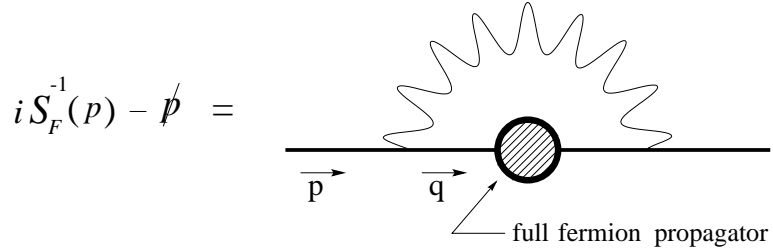}
  \end{center}
\caption{A graphical expression of the SD equation 
         in the (improved) ladder approximation.}
\label{fig:SDeq}
\end{figure}

Schwinger-Dyson (SD) equation is a powerful tool to study the 
dynamical generation of the fermion mass directly from QCD
(for reviews, see, e.g., Refs.~\cite{Kugo:review,Miransky:book}).
The SD equation for the full fermion propagator 
$\ i S_{F}^{-1} = A(p^2) \fsl{p} - B(p^2)\ $ in the improved 
ladder approximation~\cite{Miransky:vj,Higashijima:1983gx} 
is given by 
(see Fig.~\ref{fig:SDeq} for a graphical expression)
\begin{equation}
  i S_F^{-1}(p) - \fsl{p}\ =\  C_2\int \frac{d^4 q}{i (2 \pi)^4}\ 
  \bar{g}^2(p,q)\ \frac{1}{(p-q)^2}
  \left( g_{\mu \nu} - \frac{(p-q)_\mu(p-q)_\nu}{(p-q)^2} \right)
  \gamma^\mu \ i S_F(q) \ \gamma^\nu ,
\label{eq:improved_ladder_SD_0}
\end{equation}
where $\ C_2 \left(  = \frac{N^2_c - 1}{2 N_c} \right) \ $
 is the second casimir invariant, and $\ \bar{g}(p,q)\ $ is 
the  running coupling.
The Landau gauge is adopted for the gauge boson propagator.
The SD equation provides coupled equations for
two functions $A$ and $B$ in 
the full fermion propagator $S_{F}$.
When we use a simple ansatz for the 
running coupling, $ \bar{g}^2(p,q) =\bar{g}^2(\max (p_E^2, q_E^2))
$~\cite{Miransky:vj,Higashijima:1983gx},
with $(p_E^2, q_E^2)$ being the Euclidean momenta,
we can carry out the angular integration and get $A(p^2)\equiv 1$
in the Landau gauge. Then the SD equation becomes a self-consistent 
equation for
the mass function $\Sigma(p^2) \equiv B(p^2)$. The resultant
asymptotic behavior of the dynamical mass 
$\Sigma(p^2)$ 
is shown to coincide
with that obtained by the operator product expansion 
technique~\cite{Kugo:review,Miransky:book}. 

However, 
it was shown in Ref.~\cite{Kugo:1992pr} that the axial Ward-Takahashi 
identity is violated in the improved ladder approximation unless the
gluon momentum is used as the argument of the running coupling
as $\bar{g}^2( (p_E-q_E)^2 )$.
In this choice
we cannot
carry out the angle integration 
analytically since the running coupling depends on the 
angle factor 
$\cos{\theta} = p_E \cdot q_E / \vert p_E \vert \vert q_E \vert $. 
Furthermore, we would need to introduce a nonlocal gauge 
fixing~\cite{Kugo:1992pr} to preserve the condition $A=1$.

In Ref.~\cite{Aoki:1990aq}, however, it was shown that 
an angle averaged form
$\bar{g}^2(p_E^2+q_E^2)$ 
gives a good approximation.
Then, in the present analysis
we take the argument of the running coupling as
\begin{equation}
  \bar{g}^2(p_E,q_E) \ \Rightarrow\  \bar{g}^2(p_E^2+q_E^2) .
\label{eq:approx_of_argument}
\end{equation}
After applying this angle approximation and carrying out the angular
integration, 
we can show (see, e.g., 
Refs.~\cite{Leung-Love-Bardeen,Appelquist:1988yc,
Kugo:review,Miransky:book})
that $A$ always satisfies $A(p^2)=1$ in the Landau gauge.
Then the SD equation becomes 
\begin{equation}
  \Sigma(x) = C_2\  \frac{3}{16 \pi^2} \int dy\  
  \frac{y\  \Sigma(y)}{y + \Sigma^2(y)} 
  \ \frac{\bar{g}^2(x+y)}{\max(x,y)} \ ,
\label{eq:improved_ladder_SD}
\end{equation}
where $\ x = p_E^2 \ $ and $\ y = q_E^2 \ $.
Although the choice of arguments in Eq.~(\ref{eq:approx_of_argument})
explicitly breaks 
the chiral symmetry as mentioned above, it will be shown later 
that the magnitude of the breaking is negligible.

\subsection{Running coupling in large $N_f$ QCD}
\label{sec:large_Nf}

In QCD
with $N_f$ flavors of massless quarks,
the renormalization group equation (RGE) for the running coupling
$\alpha(\mu)  \ \Big(= \frac{\bar{g}^2(\mu)}{4 \pi}\,\Big)$ 
in the two-loop approximation is given by 
\begin{equation}
  \mu \frac{d}{d \mu} \alpha(\mu) 
  = -b \alpha^2(\mu) - c \alpha^3(\mu) ,
\label{eq:RGE_for_alpha}
\end{equation}
where
\begin{equation}
  b = \frac{1}{6 \pi} \left( 11 N_c - 2 N_f \right)\ ,\quad
  c = \frac{1}{24 \pi^2} \left( 34 N_c^2 - 10 N_c N_f
      - 3 \frac{N_c^2 - 1}{N_c} N_f \right) \ .
\end{equation}
{}From the above beta function we can easily see that,
when $b>0$ and $c<0$, i.e.,
$N_f$ takes a value in the range of 
$  N_f^* < N_f < \frac{11}{2} N_c $ 
($N_f^\ast \simeq 8.05$ for $N_c = 3$), 
the theory is asymptotically free and 
the beta function
has 
a zero, corresponding to
an infrared stable fixed 
point~\cite{Banks:nn,Appelquist:1996dq},
at
\begin{equation}
  \alpha_\ast = - \ \frac{\ b\ }{\ c\ } \ .
\label{eq:alpha_IR}
\end{equation}

\begin{figure}
  \begin{center}
    \includegraphics[height=6cm]{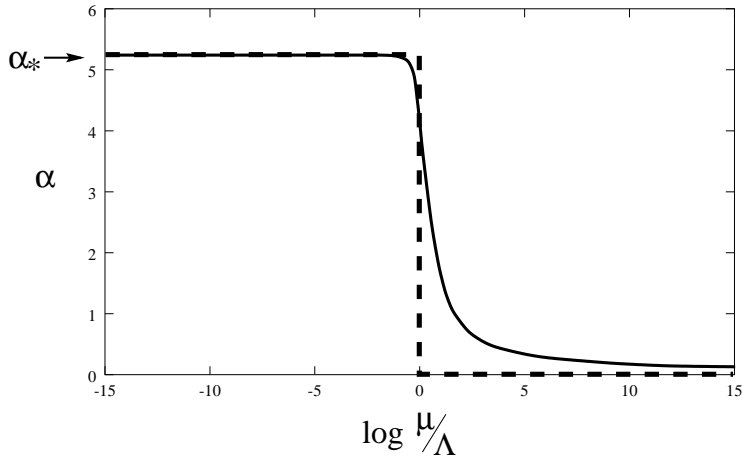}
  \end{center}
\caption{
Two-loop
running coupling (solid line) compared with the approximate
form in Eq.~(\ref{eq:step}) (dashed line)
for $N_f = 9$.
}
\label{fig:step}
\end{figure}

Existence of the infrared fixed point 
implies that
the running coupling takes a finite value 
even in the low energy region.
Actually,
the solution of the two loop
RGE in Eq.~(\ref{eq:RGE_for_alpha}) can be explicitly 
written~\cite{Gardi,ExplicitSolution} 
in all the energy region
as 
\begin{equation}
  \alpha(\mu) = \alpha_\ast \left[\ W(\mu^{\, b \alpha_\ast} 
  / e \Lambda^{ b \alpha_\ast} ) + 1\ \right]^{-1},
\label{Lan W}
\end{equation}
where $W(x) = F^{-1}(x)$ with $F(x) = x e^x$ is the Lambert $W$
function, and 
$\Lambda$ is a renormalization group invariant
scale defined by~\cite{Appelquist:1996dq}
\begin{equation}
  \Lambda \ \equiv\  \mu \ \exp \left[ - \frac{1}{b\  \alpha_\ast}
         \log \left( \frac{\alpha_\ast - \alpha(\mu)}{\alpha(\mu)}
         \right)
        - \frac{1}{b\  \alpha(\mu)}
         \right] .
\label{eq:Lambda}
\end{equation}
We note that, in the present analysis, we fix the value of
$\Lambda$ to compare the theories with a different number of flavors, 
and that we have no adjustable parameters in the
running coupling in Eq.~(\ref{Lan W}), accordingly 
(see discussion below).
We show 
an example of $\alpha(\mu)$ for $N_f = 9$ 
by the solid line in Fig.~\ref{fig:step}.

The fact that the running coupling is expressed by a
certain function as in Eq.~(\ref{Lan W}) implies that,
in the case of large $N_f$ QCD,
{\it we do not need to introduce any infrared regularizations}
such as the ones adopted in 
Refs.~\cite{Higashijima:1983gx,Miransky:vj,Aoki:1990eq}
for studying real-life QCD with small $N_f$
in which
the infrared regularization parameter must be chosen 
in such a way that
the running coupling in the infrared region becomes larger
than the critical value $\alpha_{\rm cr} = \pi/4$
for realizing the dynamical chiral symmetry 
breaking~\cite{Higashijima:1983gx}.
The running coupling in large $N_f$ QCD takes a certain
value in the infrared region for given $N_f$, so that
we can definitely determine, within the framework of the SD equation,
whether or not the dynamical chiral symmetry breaking is realized.
Actually, the value of 
$\alpha_\ast$ decreases 
monotonically with increasing $N_f$,
and
the chiral symmetry restores when $N_f$ 
becomes large enough.
In Refs.~\cite{Appelquist:1996dq,Appelquist:1998rb}, 
it was shown that the phase transition occurs at 
 $N_f^{\rm crit} \simeq  11.9$ for $N_c = 3$
(corresponding to $\alpha_\ast = \alpha_{\rm cr} = \pi/4$).

In order to reduce the task of numerical calculations 
in solving the HBS equation, 
we modify the shape of the running coupling.
Since the dynamics in the infrared region
governs the chiral symmetry breaking,
we adopt the following approximation for
the running 
coupling ~\cite{Miransky:1996pd,Appelquist:1998rb}:
\begin{equation}
  \frac{\bar{g}^2(x+y)}{4\pi} =
  \alpha_\ast\ \theta(\Lambda^2 - (x+y) ) \ .
\label{eq:step}
\end{equation}
In this approximation the coupling takes the constant value
$\alpha_\ast$ (the value at the IR fixed point) below the scale
$\Lambda$ and entirely vanishes in the energy region above this scale.
The dashed line In Fig.~\ref{fig:step} represents the approximated 
form of the running coupling for $N_f=9$.

\subsection{Numerical solution for the SD equation}
\label{sec:numerical_SD}

In this subsection we briefly explain how we solve the SD equation
numerically.

\begin{figure}
  \begin{center}
    \includegraphics[height=6cm]{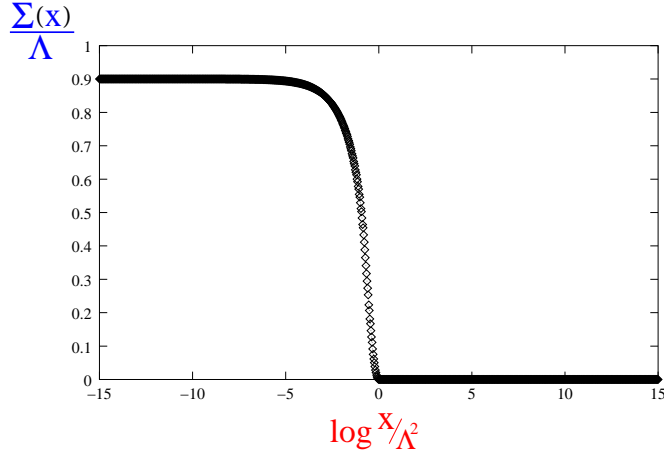}
  \end{center}
\caption[]{
A solution
of the discretized SD equation 
in Eq.~(\ref{eq:discretized_largeNf_SD_1})
for $N_f = 9$.
}
\label{fig:mass_function}
\end{figure}

We first introduce the infrared
(IR) cutoff $\lambda_{SD}$ and ultraviolet (UV) cutoff 
$\Lambda_{SD}$ as
\begin{equation}
\Lambda^2 \ e^{\lambda_{SD}/\Lambda} \ \le\  x \, , \ y \ \le \ 
\Lambda^2 \ e^{\Lambda_{SD}/\Lambda}
\ .
\end{equation}
Then, we discretize the momentum variable $x$ and  
$y$ into $N_{SD}$ points as
\begin{equation}
  x_i =  \Lambda^2 \ \exp \Big[ \lambda_{SD}/\Lambda +  
                       D_{SD}\cdot i\,
                       \Big] \ , \ \ \ 
  \Big( i = 0, 1, 2, \cdots, (N_{SD}-1) \Big) \ ,
\end{equation}
where
\begin{equation}
  D_{SD} = \frac{\ ( \Lambda_{SD} - \lambda_{SD} )/\Lambda\ }{N_{SD} - 1} \ .
\end{equation}
Accordingly, the integration over $y$ is replaced with a summation as
\begin{equation}
\int d y \ \Rightarrow \ D_{SD} \sum_{j} y_j \ .
\end{equation}
Then, the SD equation in Eq.~(\ref{eq:improved_ladder_SD}) with the
running coupling in Eq.~(\ref{eq:step}) is rewritten as
\begin{equation}
  \Sigma(x_i) \ =  \ \frac{1}{4 \pi^2}  
         \ D_{SD}\sum_j \  
         \bar g^2(x_i + y_j) \
         \frac{y_j^2}{\max(x_i,\ y_j)}
         \ \frac{\Sigma(y_j)}
         {y_j + \Sigma^2(y_j)} .
\label{eq:discretized_largeNf_SD_1}
\end{equation}
This discretized version of the
SD equation is solved 
by the recursion relation:
\begin{equation}
  \Sigma_{(n+1)}(x_i) =  \frac{1}{4 \pi^2}  
          D_{SD}\sum_j 
          \bar g^2(x_i + y_j) \
          \frac{y_j^2}{\max(x_i,\ y_j)}
          \ \frac{\Sigma_{(n)}(y_j)}
          {y_j + \Sigma_{(n)}^2(y_j)} \ .
\label{eq:discretized_largeNf_SD_2}
\end{equation}
Starting from 
a suitable initial condition (we choose
$\Sigma_{(0)}(x_i) = 1 $),
we update the mass function by the above recursion relation.
Then, we stop the iteration when the convergence condition 
\begin{equation}
  D_{SD}\sum_i
       \frac{x^2_i}{16 \pi^2} \,
       \Big[\,\Sigma_{(n+1)}(x_i) 
               - \Sigma_{(n)}(x_i)\,\Big]^2 
  \ <\  \varepsilon^2 \Lambda^6
\end{equation}
is satisfied for sufficiently small $\varepsilon$,
and
regard this $\Sigma_{(n)}$ as a solution of 
Eq.~(\ref{eq:discretized_largeNf_SD_1}).
In Fig.~\ref{fig:mass_function}, we show the numerical 
solution for the mass function $\Sigma(x)$. 
Here, we took $N_f = 9\ $ ($\alpha_\ast\simeq5.2$)
as an example and used 
the following parameters:
\begin{equation}
  \Lambda_{SD}/\Lambda = + 15 \ \ ,\ \ 
  \lambda_{SD}/\Lambda =  - 15 \ \ ,\ \ 
  N_{SD}  = 1000 \ \ ,\ \ 
  \varepsilon = 10^{-15} .
\end{equation}

\begin{figure}
  \begin{center}
    \includegraphics[height=7cm]{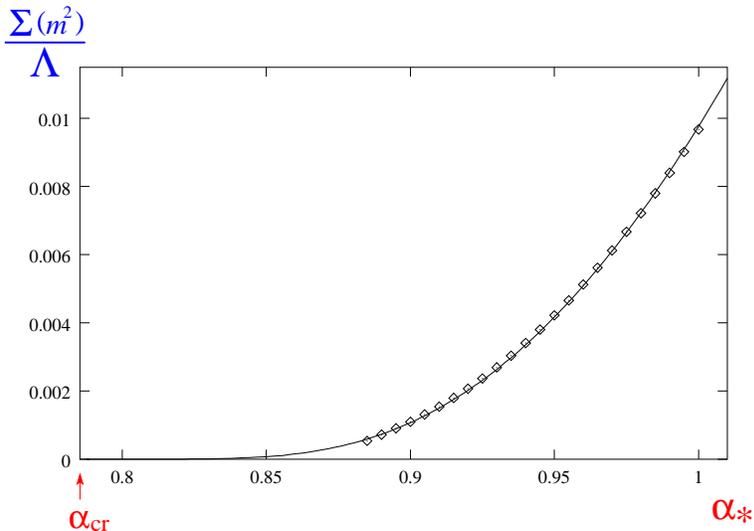}
  \end{center}
\caption{
  Numerical solutions
  of $\Sigma(x=m^2)$ 
  for several values of $\alpha_\ast$
  (indicated by $\Diamond$). 
  Solid line shows the function in Eq.~(\ref{eq:fitting})
  with the best fitted value $d = 4.0$.
}
\label{fig:Sigma_IR}
\end{figure}

Now, let us study
the critical
behavior of the fermion mass 
as $N_f$ is varied.
Note that
we can use $\alpha_\ast$ instead of $N_f$ as an input parameter, 
because once we choose a value of $N_f$, the value of $\alpha_\ast$ 
is uniquely determined from Eq.~(\ref{eq:alpha_IR}).
For example, 
 $\alpha_\ast = 1$ implies $N_f = 11.42$ and 
 $\alpha_\ast = \alpha_{\rm cr}$ implies $N_f = 11.91$.
We solve 
Eq.~(\ref{eq:discretized_largeNf_SD_1}) for various 
values of $\alpha_\ast$
and plot the values of $\ \Sigma(m^2)\ $ 
in Fig.~\ref{fig:Sigma_IR}. 
Here, $\ m\ $ represents the dynamical mass defied by 
$\ m = \Sigma(m^2)\ $.
It should be noticed that $m$ is defined in the space-like region
which does not represent the pole mass of fermion.  As we will 
show in section~\ref{sec:complex_SD}, 
the present fermion propagator does not have
any poles and then there are no pole masses of fermion.

We compare this result with the analytic 
solution~\cite{Miransky:1996pd,Appelquist:1998rb}:
\begin{eqnarray}
 \Sigma(m^2) &\approx& \Lambda \ \exp 
       \left(
            - \frac{\pi}{\sqrt{\frac{\alpha\ast}{\alpha_{\rm cr}}-1\ }}
       \right) 
    \hspace{1.14cm}  \mbox{for}\ \  \alpha_\ast > \alpha_{\rm cr} \ .
\label{eq:Sigma_m}
\end{eqnarray}
In the above form there is an ambiguity in the prefactor.
Then, we introduce
the function
\begin{equation}
  h(\alpha_\ast) = d \,\Lambda \,\exp 
       \left(
            - \frac{\pi}{\sqrt{\frac{\alpha\ast}{\alpha_{\rm cr}}-1\ }}
       \right)
   \ ,
\label{eq:fitting}
\end{equation}
and fit the value of the pre-factor $d$ by minimizing
\begin{equation}
\sum_{\alpha_\ast}
\left\vert\,
  \Sigma (x=m^2;\alpha_\ast) - h(\alpha_\ast) \,
\right\vert^2
\ ,
\end{equation}
in the range of $\alpha_\ast \in [0.885:1]$.
The resultant best fitted value of $d$ is
\begin{equation}
  d \simeq 4.0 \ .
\end{equation}
We plot the function in Eq.~(\ref{eq:fitting}) with the best fitted
value $d = 4.0$ in Fig.~\ref{fig:Sigma_IR} (solid line).
This clearly shows that the $\alpha_\ast$-dependence of the resultant
$\Sigma(m^2)$ from our numerical calculation is consistent with the
analytic result:
The dynamical mass function vanishes when $\alpha_\ast$ reaches the
critical value $\alpha_{\rm cr} = \pi/4$.
Noting that decreasing $\alpha_\ast$ corresponds to increasing
$N_f$ for fixed $N_c$ as we discussed in the previous subsection,
we see that 
the chiral symmetry restoration actually occurs at 
$N_f = N_f^{\rm crit} \simeq  
  12\,(N_c/3)$~\cite{Appelquist:1996dq,Appelquist:1998rb}.

\subsection{Pseudoscalar meson decay constant in large   $N_f$ QCD}
\label{sec:f_pi}

\begin{figure}
  \begin{center}
    \includegraphics[height=7cm]{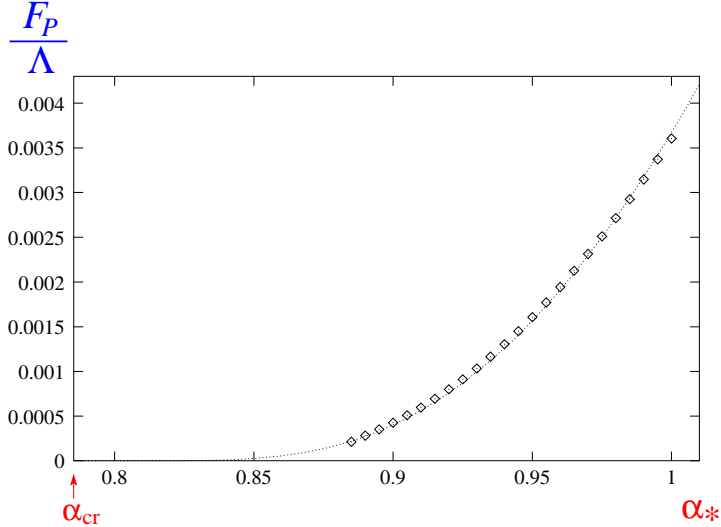}
  \end{center}
\caption{
  Values of $F_P$ calculated from
  the Pagels-Stokar formula for several values of $\alpha_\ast$
  (indicated by $\Diamond$).
  The dotted line shows the function of the form in 
  Eq.~(\ref{eq:fitting}) with the best fitted value 
  $d = 1.5$.
}
\label{fig:F_pi}
\end{figure}

So far
we have solved the SD equation and obtained the mass functions 
for the various values of $\alpha_\ast$.
Now we can calculate the pseudoscalar meson 
decay constant $F_P$ at each $\alpha_\ast$ by using 
the Pagels-Stokar formula \cite{Pagels:hd}:
\begin{equation}
  F_P^2 \ = \ \frac{N_c}{4\pi^2} \int dx \ 
       \frac{\ x\ \Sigma^2(x) \ -\  \frac{\ x^2}{4} \ \frac{d}{dx}
       \left[\  \Sigma^2(x) \ \right]\ }
       {\left(\  x \ +\  \Sigma^2(x) \ \right)^2} .
\end{equation} 
In Fig.~\ref{fig:F_pi}, we plot the values of
$F_P$
for several values of 
$\alpha_\ast$
(indicated by $\Diamond$).
To study the critical behavior of the pseudoscalar meson 
decay constant we use the
function of the form in Eq.~(\ref{eq:fitting}) and
fit the value of $d$ by minimizing
\begin{equation}
\sum_{\alpha_\ast}
\left\vert
  F_P(\alpha_\ast) - h(\alpha_\ast)
\right\vert^2
\ ,
\end{equation}
for $\alpha_\ast \in [0.885:1]$.
The resultant best fitted value of $d$ is 
\begin{equation}
  d \simeq 1.5 \ (\equiv d_{F_P}) \ .
\label{eq:d_fpi}
\end{equation}
We plot the fitting function with $d = 1.5$ in Fig.~\ref{fig:F_pi}
(dotted line).
This shows that the results of the numerical calculations 
for $F_P$ are well fitted by
the function of the form in Eq.~(\ref{eq:fitting}), and that
the pseudoscalar meson 
decay constant has the same critical behavior as the mass
function has.

\subsection{Fermion-antifermion pair condensate}
\label{sec:pair_condensate}
In this subsection, we calculate the fermion-antifermion pair 
condensate $\langle \bar\psi \psi \rangle$ in large $N_f$ QCD, 
and show that the system in the present analysis has the large 
anomalous dimension $\gamma_m$ for the operator $\bar\psi \psi$.
We also show that the values of $\gamma_m$ 
are not affected so much by the approximation for the running coupling 
used in the present analysis.

The condensate
$\langle \bar\psi \psi \rangle$ 
is calculated from the following equation:
\begin{equation}
  \langle \bar\psi \psi \rangle_{\Lambda_{UV}}
       \ =\  - \frac{N_c}{4\pi^2} 
       \int_{0}^{\Lambda_{UV}^2} dx \ 
       \frac{x\ \Sigma(x)}
       {\ x + \Sigma^2(x)\ } ,
\label{eq:q-bar-q}
\end{equation}
where $\Sigma(x)$ is the mass function obtained from
the SD equation and
$\Lambda_{UV}$ represents UV cutoff introduced to regularize the
UV divergence.
In the improved ladder approximation, the high-energy behavior of the
mass function is consistent with that derived using the operator
product expansion (OPE).  
The chiral condensate calculated using the mass function was shown to
obey the renormalization group equation derived with the OPE
(see, e.g., Refs.~\cite{Kugo:review,Miransky:book}).
Then, as was adopted in 
Refs.\cite{Appelquist:1996dq,Miransky:1996pd,Appelquist:1998rb}, 
we identify the condensate,
which is calculated with UV cutoff $\Lambda_{UV}$, with that 
renormalized at the scale $\Lambda_{UV}$ in QCD.
\footnote{When the condensate is calculated 
using the approximated running coupling 
defined by Eq.~(\ref{eq:step}),
the integration in Eq.~(\ref{eq:q-bar-q}) is effectively 
cut off at the scale of $\Lambda$ due to the truncation of 
the running coupling for any values of $\Lambda_{UV} > \Lambda$.
(See Fig.~\ref{fig:mass_function} : $\Sigma(x) = 0$ for $x >
\Lambda^2$.) 
}

We expect that infrared dynamics in large $N_f$ QCD is similar
to that of strong coupling QED 
or walking gauge theories~\cite{Yamawaki:1985zg} 
since the running coupling in large $N_f$ QCD is well approximated 
by the constant coupling (see Fig.~\ref{fig:step})
~\cite{Appelquist:1998rb}.
Then, we also expect that the value of the anomalous dimension 
in large $N_f$ QCD becomes $\gamma_m \simeq 1$ since
the walking gauge theories 
have $\gamma_m \simeq 1$
~\cite{Yamawaki:1985zg}.

When a considering system has the anomalous dimension $\gamma_m$, 
scaling properties of $F_P$ and $- \langle \bar\psi \psi \rangle$ 
with respect to $\alpha_\ast$ 
near the critical point are expressed as 
follows~\cite{Yamawaki:1985zg} :
\begin{equation}
  F_P \sim m ,
\label{eq:scaling of fpi}
\end{equation}
\begin{equation}
  - \langle \bar\psi \psi \rangle \sim m^{3 - \gamma_m} ,
\label{eq:scaling of <qq>}
\end{equation}
where $m$ represents the dynamical fermion mass.
These equations mean that the relation between 
$\langle \bar\psi \psi \rangle$ and $F_P$ can be written as 
\begin{equation}
  - \langle \bar\psi \psi \rangle = c \cdot F_P^{3 - \gamma_m},
\end{equation}
where $c$ is a certain positive constant.
From this equation, we can express the anomalous
dimension as 
\begin{equation}
  \gamma_m \ =\ \gamma'_m \ -\  \varepsilon,
\end{equation}
where
\begin{eqnarray}
  \gamma'_m &=& 3 - \frac{\log ( - \langle \bar\psi \psi \rangle )}
                      {\log F_P}, \label{eq:gamma'_m}\\
  \varepsilon &=&  \frac{\log c}{- \log F_P}.
\end{eqnarray}
Here, we note that $\gamma'_m$ approaches $\gamma_m$ for $\alpha_\ast
\rightarrow \alpha_{\rm cr}$ since $F_P$ becomes small, 
i.e., $\ (-\log F_P) \gg 1$, 
near the critical point (see Fig.~\ref{fig:F_pi}):
\begin{equation}
  \varepsilon \rightarrow 0 \ \ \ \ \ \ 
  \mbox{for} \ \ \alpha_\ast \rightarrow  \alpha_{\rm cr}.
\end{equation}

\begin{figure}
  \begin{center}
    \includegraphics[height=7cm]{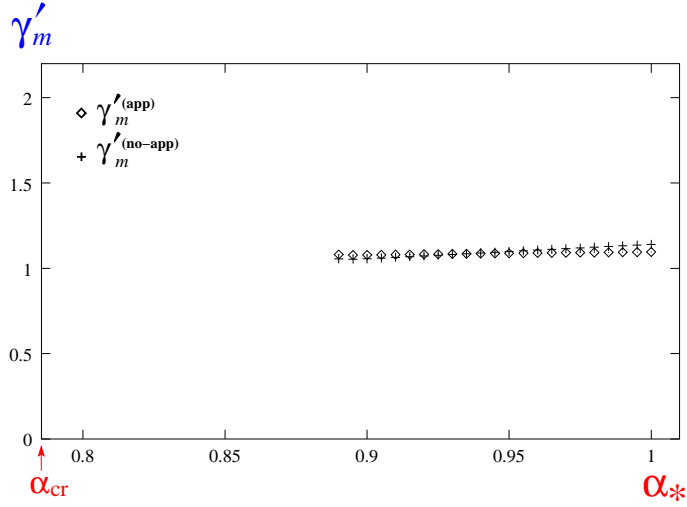}
  \end{center}
\caption{Values of $\gamma'_m$ for several values of $\alpha_\ast$.
         $\Diamond$ and $+$ represent the values of $\gamma'_m$ 
         calculated from Eq.~(\ref{eq:gamma'_m}) with and 
         without approximation for the running coupling.
         (We call them $\ {\gamma'}_m^{{\bf (app)}}\ $ and 
         $\ {\gamma'}_m^{{\bf (no-app)}}\ $ respectively.)
}
\label{fig:q-bar-q_over_fpi2}
\end{figure}

In Fig.~\ref{fig:q-bar-q_over_fpi2}, we plot the values of 
$\gamma'_m$ for several values of $\alpha_\ast$ 
as an estimation of the anomalous dimension.
The data indicated by $\Diamond$ in Fig.~\ref{fig:q-bar-q_over_fpi2} 
is obtained with the approximated 
running coupling (dashed line in Fig.~\ref{fig:step}) 
in the SD equation. 
(We call this kind of data ${\gamma'}_m^{\bf (app)}$.)
On the other hand, the data indicated by $+$  
is the result from the calculation 
with the two-loop running coupling given in Eq.~(\ref{Lan W}).
(We call this kind of data 
$\ {\gamma}_m^{{\bf (no-app)}}\ $.)
\footnote{The reason why we introduced the approximated 
running coupling (\ref{eq:step}) in the present analysis 
is to reduce the task of numerical calculations 
in solving the HBS equations.
As for the SD equation, 
we can easily solve it numerically
with the two-loop running coupling given in Eq.~(\ref{Lan W}).

Since we have to compare 
$\ {\gamma'}_m^{{\bf (app)}}\ $ 
and  
$\ {\gamma'}_m^{{\bf (no-app)}}\ $ 
at the same energy scale, we have lowered the scale of 
$\langle \bar\psi \psi \rangle^{{\bf (no-app)}}\ $ 
from $\Lambda_{UV}$ to $\Lambda$ by the following two-loop 
renormalization group equation : 
\[
  \langle \bar\psi \psi \rangle_{\Lambda} \ =\  
  \langle \bar\psi \psi \rangle_{\Lambda_{UV}} 
  \left[ \frac{ \alpha(\Lambda_{UV}) }{ \alpha(\Lambda) }
  \right]^{\frac{ \gamma_m^{(0)} }{ 4 \pi b }} 
  \left[ 1 + \left( \frac{ \gamma_m^{(1)} }{ 4 \pi b } 
       - \frac{ 2 \gamma_m^{(0)} }{ b^2 } \right) 
    \frac{ \alpha(\Lambda) - \alpha(\Lambda_{UV}) }{ 4 \pi }
  \right],
\]
where, 
\[
  \gamma_m^{(0)} = 6 C_2 \ ,\ \ \ 
  \gamma_m^{(1)} = C_2 \left( 3 C_2 + \frac{97}{3} N_c - 
     \frac{10}{3} N_f 
     \right).\nonumber
\]
}
From these results, we conclude that large $N_f$ QCD with 
two-loop running coupling as well as with approximated 
running coupling actually possesses 
\footnote{
From the values of $c$ obtained by fitting to the data 
of $\langle \bar\psi \psi \rangle$, 
we find $\varepsilon = 0.04 \sim 0.06$ for the approximated running
coupling, and $\varepsilon = 0.16 \sim 0.25$ 
for the two-loop running coupling.
}
\begin{equation}
\gamma_m \simeq \gamma'_m \simeq 1.
\end{equation} 

Moreover, Fig.~\ref{fig:q-bar-q_over_fpi2} shows that 
the data of 
$\ {\gamma'}_m^{{\bf (app)}}\ $ 
is in good agreement with that of 
$\ {\gamma'}_m^{{\bf (no-app)}}\ $, which implies   
that the approximation 
for the running coupling used in the present analysis works well.
We also expect that the approximation does not affect the results 
so much when we calculate the HBS equations for the bound states.

\section{SD equation on the complex plane}
\label{sec:complex_SD}

As we
will discuss in section~\ref{sec:HBSeq}, we need the mass 
function for
complex arguments
when we solve the HBS equation 
for the massive bound state. 
In this section,
we first introduce
the SD equation 
for the complex argument following 
Ref.~\cite{Kugo-Yoshida} (see also Ref.~\cite{Harada:1995nx}), 
and then
solve it 
in the case of large $N_f$ QCD.

The SD equation 
for the complex argument
is expressed as~\cite{Kugo-Yoshida}
\begin{equation}
  \Sigma (x) = C_2 \frac{3}{16\pi^2}
               \left[ \int_{C(0,x)} dy \frac{y}{x}
                  + \int_{C(x,\infty)} dy  \right]
               \frac{\bar{g}^2(x+y)\,
               \Sigma (y)}{y + {\Sigma^2 (y)}} \ ,
\label{eq:complex_SD}
\end{equation}
where $C(a,b)$ is the integral path on the complex plane.
Here, we took
the same argument of the running coupling
as that in
Eq.~(\ref{eq:approx_of_argument}), and 
carried out the angle integration.
Note that the integral path $C(a,b)$ must be taken so as to avoid the
branch cut appearing in the integral.

We first study the structure of the running
coupling appearing in the SD equation (\ref{eq:complex_SD})
to clarify the branch cut.
In the improved ladder approximation it is essential to use
the running coupling determined from the $\beta$-function 
in the high energy (space-like) region 
for consistency with perturbative QCD.
In QCD with small $N_f$, however, the running coupling obtained
from the perturbative $\beta$-function diverges at some infrared
scale, $\Lambda_{\rm QCD}$.
In the ordinary SD equation in the space-like region,
the infrared singularity is avoided by introducing infrared
regularization such as the so-called Higashijima-Miransky 
approximation~\cite{Higashijima:1983gx,Miransky:vj}
and its extension as in Ref.~\cite{Aoki:1990eq}.
However, 
since the running coupling in Eq.~(\ref{eq:complex_SD}) is
a complex function which has the complex argument,
we need an extension with analyticity satisfied.
Several works such as
in Refs.~\cite{Kugo-Mitchard-Yoshida,Maris-Roberts-Tandy}
proposed models of running coupling which are consistent
with perturbative QCD in the high energy region.
But they still have branch cuts on the complex plane,
and it is a burdensome task to evade all the branch cuts
by carefully selecting the integral path in 
Eq.~(\ref{eq:complex_SD}).
One way to avoide such a complication might be to
give up the consistency with perturbative QCD
and use models of running coupling with analyticity such as the
one used in Ref.~\cite{Alkofer-Watson-Weigel}.

Here we point out that the situation dramatically changes in the
large $N_f$ QCD.
In the case of large $N_f$ QCD, 
as we explained in subsection~\ref{sec:large_Nf},
the running coupling, as well as
the two-loop $\beta$-function, 
is finite for any space-like momentum.
This implies that we may be able to construct the 
running coupling 
by analytic continuation using the $\beta$-function.
Actually,
an explicit solution of the two-loop 
renormalization group equation (RGE) can be written 
in terms of the Lambert $W$ 
function~\cite{Appelquist:1998rb,Gardi}.
When $N_f$ is close to $N_f^{\rm crit}$, 
the solution of the RGE has 
no singularity on the complex plane except for the time-like 
axis~\cite{Gardi}.

\begin{figure}
  \begin{center}
    \includegraphics[height=6cm]{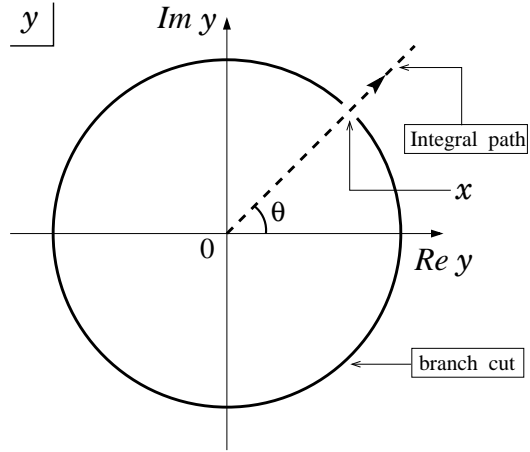}
  \end{center}
\caption[]{Integral path
  of the SD equation (\ref{eq:complex_SD}).
  Here, the branch cut appears from the four-dimensional angle
  integration.
  }
\label{fig:IntegralPath}
\end{figure}

As a result,
for general complex $x$ except on the time-like axis ($x<0$),
we can take the integral path $C(a,b)$ in such a way 
that it just avoids the branch cut coming from the angle integration. 
In Fig.~\ref{fig:IntegralPath} we show the branch cut together
with a simple choice of the integral 
path~\cite{Kugo-Yoshida}.
We stress again that the reason why
we can take this simple integral path is that
the running coupling has no singularity 
on the complex plane except for the time-like axis.

\begin{figure}
  \begin{center}
    \includegraphics[height=6cm]{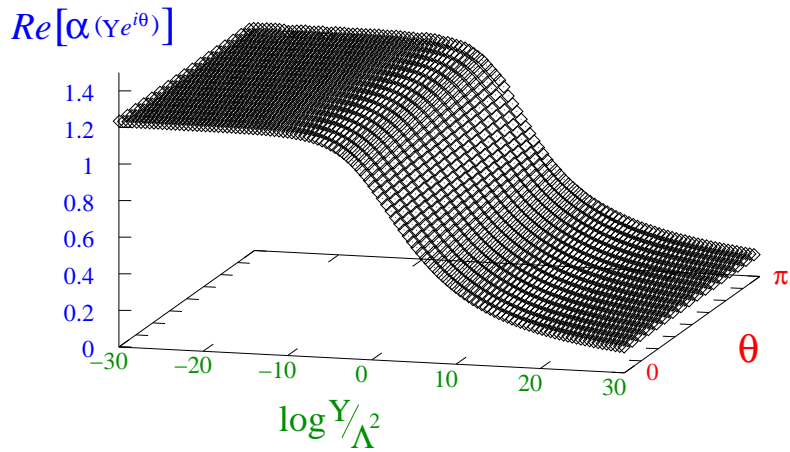}
  \end{center}
\caption{Real part of the two-loop running coupling
for $N_f=11$ on the complex plane obtained 
by the analytic continuation from the running coupling
on the real axis 
(we use the Cauchy-Rieman relation). 
The complex argument of $\alpha$ is expressed as 
$y = Y e^{i \theta}$, where $Y$ and $\theta$ are real.
Note that $y$ is in the space-like region for $\theta=0$ 
and in the time-like region for $\theta=\pi$.
}
\label{fig:analytic_cont}
\end{figure}

For solving the SD equation on the complex plane,
we here study the explicit form of the running coupling.
In Fig.~\ref{fig:analytic_cont} we show the real part of 
the running coupling 
on the complex plane which is obtained by performing the analytic 
continuation from the running coupling on the real axis determined 
from the two-loop $\beta$-function.
This figure shows that $\mbox{Re} \alpha \simeq \alpha_\ast$,
i.e., $\mbox{Im}\alpha \simeq 0$,
in 
the range of $Y = \vert y \vert < \Lambda^2$, 
and that $\mbox{Re} \alpha \ll \alpha_\ast$ in the 
range of $Y > \Lambda^2$.
Thus 
we take the following approximation for the running coupling
on the complex plane:
\begin{equation}
  \alpha (y) \ =\  \alpha_\ast\ \theta(\Lambda^2 - Y) 
\ ,
\label{coupling for complex}
\end{equation}
which is smoothly connected to the approximation adopted in
Eq.~(\ref{eq:step}) for the running coupling on the space-like
axis.

Now, let us solve the SD equation (\ref{eq:complex_SD}) to obtain the
mass function for complex variable $x$.
Along the
integral path shown in Fig.~\ref{fig:IntegralPath}, 
the variables $x$ and $y$ are parametrized as
\begin{equation}
  x = X e^{i \theta},\ \  y = Y e^{i \theta},
\end{equation}
where $X$, $Y$, and $\theta$ are real.
Then the SD equation (\ref{eq:complex_SD}) is rewritten as
\begin{equation}
  \Sigma (X e^{i\theta}) = C_2 \frac{3}{16\pi^2}
               \left[ \int_0^X dY \frac{Y}{X}
                  + \int_X^\infty dY  \right]
               \frac{\bar{g}^2( X+Y )\,
  \Sigma (Y e^{i\theta})}{Y + e^{-i\theta}\,\Sigma^2 (Y e^{i\theta})}
\ .
\label{eq:complex_SD 2}
\end{equation}
{}From this we can easily see that the solution is expressed as
\begin{equation}
  \Sigma (x) = e^{i \frac{\theta}{2} }\,\bar\Sigma (X)  \ ,
\label{eq:Sigma asum}
\end{equation}
where $\bar\Sigma (X)$ is real and satisfies the original
SD equation on the real axis:
\begin{equation}
  \bar\Sigma (X) =  C_2 \frac{3}{16\pi^2}
                    \left[ \int_{0}^{X} dY \frac{Y}{X}
                  + \int_{X}^{\infty} dY  \right]
                    \frac{\bar{g}^2( X+Y )\,
                    \bar\Sigma (Y)}{Y + {\bar\Sigma (Y)}^2} 
\ .
 \label{eq:real SD}
\end{equation}
Note that the fermion propagator $S_F$ does not have any poles 
since the kinetic part $x$ and the mass part $\Sigma^2(x)$ have
the same phases as
$x + \Sigma^2(x) = e^{i\theta} (X + \bar{\Sigma}^2(X))$
[see Eq.~(\ref{eq:Sigma asum})]
and the mass function in the space-like region satisfies
$X + \bar{\Sigma}^2(X) > 0 $.

We should note that the above solution in 
Eq.~(\ref{eq:Sigma asum}) 
is a double-valued 
function on the complex plane:
The variable $x=X e^{i\theta}$ can be parametrized as
$x=X e^{i(\theta+2\pi)}$ for which the solution takes
$\Sigma(x) = e^{i(\theta/2 + \pi)} \bar{\Sigma}(X)
= - e^{i\theta/2} \bar{\Sigma}(X)$.
This corresponds to the fact that the SD equation
has two solutions: When $\Sigma(x)$ is a solution, 
$-\Sigma(x)$ also satisfies the equation.
When we choose the range of $\theta$ as
$\theta \in [-\pi:\pi]$, the branch 
cut emerges on the time-like axis.
This choice is natural because the appearance of the branch cut 
in the time-like region seems consistent with the analytic 
structure of the running coupling.
We will see that this branch cut does not matter in calculating 
the bound state masses.

\section{Homogeneous Bethe-Salpeter equation}
\label{sec:HBSeq}
In this section we briefly review the 
homogeneous Bethe-Salpeter (HBS) equation for the 
bound states of quark and antiquark 
and show how to solve it numerically.

\subsection{Bethe-Salpeter amplitude}

In this paper, we consider the scalar, pseudoscalar, vector, and 
axial-vector bound states of quark and antiquark, and we write 
these bound states as $\vert S(q) \rangle$, $\vert P(q) \rangle$, 
$\vert V(q,\epsilon) \rangle$, and $\vert A(q,\epsilon) \rangle$,  
respectively.
Here, $q^\mu$ represents the momentum of the bound states and 
$\epsilon^\mu$ represents the polarization vector satisfying
$\epsilon \cdot q = 0$ and $\epsilon^2 = -1$.

Now, we introduce the Bethe-Salpeter (BS) amplitudes $\chi$ for 
the bound states of quark and anti-quark as follows : 
\begin{equation}
  \langle 0 \vert 
   \ T\ \psi_{\alpha f i}(x_+)\ \bar\psi_\beta^{f^\prime j}(x_-) 
  \ \vert S_a(q) \rangle
  \ =\  \delta_i^j
  \frac{(\lambda_a)_f^{f^\prime}}{\sqrt{2}}\, 
  e^{-iqX} \int \frac{d^4p}{(2 \pi)^4} 
e^{-ipr} [\chi_{(S)}(p;q)]_{\alpha\beta},
\end{equation}
\begin{equation}
  \langle 0 \vert 
    \ T\  \psi_{\alpha f i}(x_+)\ \bar\psi_\beta^{f^\prime j}(x_-) 
  \ \vert P_a(q) \rangle
  \ =\  \delta_i^j
  \frac{(\lambda_a)_f^{f^\prime}}{\sqrt{2}}\, 
  e^{-iqX} \int \frac{d^4p}{(2 \pi)^4} 
e^{-ipr} [\chi_{(P)}(p;q)]_{\alpha\beta},
\end{equation}
\begin{equation}
  \langle 0 \vert 
    \ T\ \psi_{\alpha f i}(x_+)\ \bar\psi_\beta^{f^\prime j}(x_-) 
  \ \vert V_a(q,\epsilon) \rangle
  \ =\ \delta_i^j
  \frac{(\lambda_a)_f^{f^\prime}}{\sqrt{2}}\, 
  e^{-iqX} \int \frac{d^4p}{(2 \pi)^4} 
e^{-ipr} [\chi_{(V)}(p;q,\epsilon)]_{\alpha\beta},
\end{equation}
\begin{equation}
  \langle 0 \vert 
    \ T\  \psi_{\alpha f i}(x_+)\ \bar\psi_\beta^{f^\prime j}(x_-) \ 
  \vert A_a(q,\epsilon) \rangle
  \ =\  \delta_i^j
  \frac{(\lambda_a)_f^{f^\prime}}{\sqrt{2}}\, 
  e^{-iqX} \int \frac{d^4p}{(2 \pi)^4} 
e^{-ipr} [\chi_{(A)}(p;q,\epsilon)]_{\alpha \beta},
\end{equation}
where $x_\pm = X \pm r/2$, 
$\lambda_a$ is the generator of $\mbox{SU}(N_f)$ normalized as
$\mbox{tr}[ \lambda_a \lambda_b ] = 2 \delta_{ab}$,
and ($\alpha$, $\beta$), ($f$, $f^\prime$), and ($i$, $j$) denote
the spinor, flavor, and color indices, respectively.

We can expand the BS amplitude $\chi$ in terms of the bispinor bases 
$\Gamma^i$ and the invariant amplitudes $\chi^i$ as follows :
\begin{equation}
\left[ \chi_{(S,P)}(p;q) \right]_{\alpha \beta} 
=\ \sum_{i=1}^{4} \left[ \Gamma^i_{(S,P)}(p;q) \right]_{\alpha \beta}
  \chi_{(S,P)}^{i} (p;q) ,
\label{eq:chi-SP}
\end{equation} 
\begin{equation}
\left[ \chi_{(V,A)}(p;q,\epsilon) \right]_{\alpha \beta} 
=\ \sum_{i=1}^{8} \left[ \Gamma^i_{(V,A)}(p;q,\epsilon) \right]_{\alpha \beta}
  \chi_{(V,A)}^{i} (p;q) .
\label{eq:chi-VA}
\end{equation}
The bispinor bases can be determined from spin, parity,  
and charge conjugation properties of the bound states.
The explicit forms of $\Gamma_{(S)}^i$, $\Gamma_{(P)}^i$,
$\Gamma_{(V)}^i$, and 
$\Gamma_{(A)}^i$ are summarized in Appendix~\ref{app:bispinor-bases}.

We take the rest frame of the bound state as a 
frame of reference: 
\begin{equation}
  q^\mu = ( M_B , 0 , 0 , 0 ) ,
\end{equation}
where $M_B$ represents the bound state mass.
After the Wick rotation, we parametrize $p^\mu$ by the real 
variables $u$ and $x$ as
\begin{equation}
  p \cdot q = i M_B u \ ,\  p^2 = - u^2 - x^2 .
\end{equation}
Consequently, the invariant amplitudes $\chi^i$
become functions 
in $u$ and $x$:
\begin{equation}
  \chi^i_{(S,P,V,A)} = \chi^i_{(S,P,V,A)}(u,x) .
\end{equation}
{}From the charge conjugation properties for the BS amplitude $\chi$ 
and the bispinor bases defined in 
Appendix~\ref{app:bispinor-bases}, 
the invariant amplitudes $\chi^i(u,x)$ are shown to 
satisfy the following relation:
\begin{equation}
  \chi^i_{(S,P,V,A)}(u,x) = \chi^i_{(S,P,V,A)}(-u,x)\ .
\label{eq:even_chi}
\end{equation}

\subsection{HBS equation}
\begin{figure}
  \begin{center}
    \includegraphics[height=3.5cm]{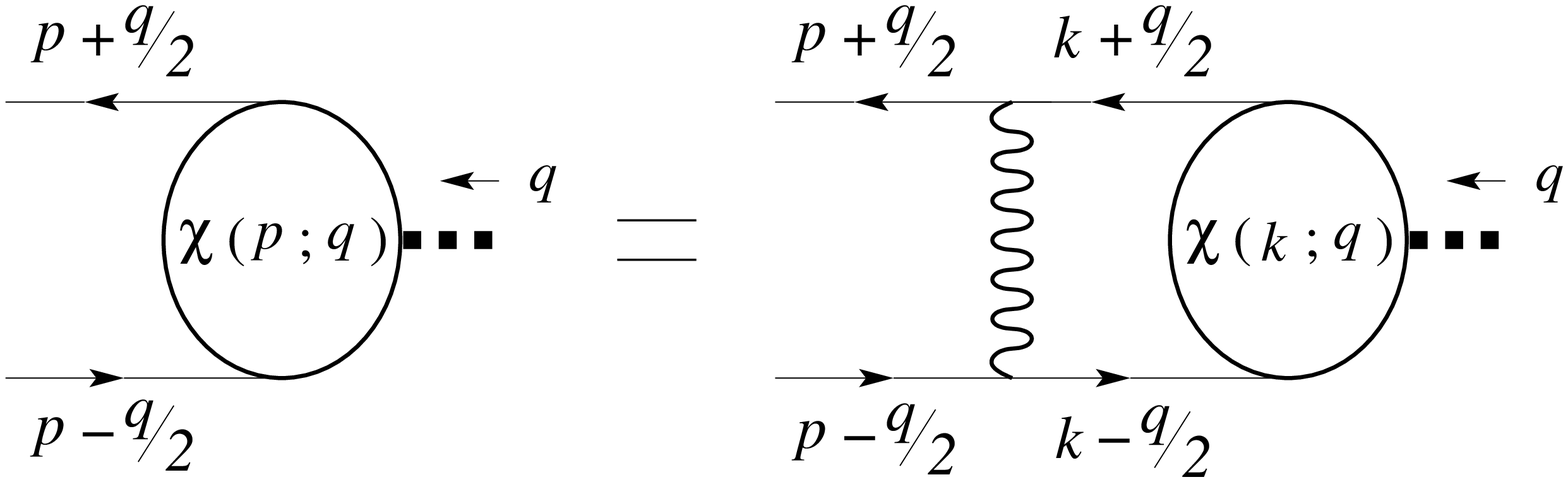}
  \end{center}
\caption{A graphical representation of the HBS equation 
in the (improved) ladder approximation}
\label{fig:HBSeq}
\end{figure}

The HBS equation is the self-consistent equation for 
the BS amplitude
(see, for a review, Ref.~\cite{Nakanishi:ph}),
and it is expressed as 
(see Fig.~\ref{fig:HBSeq})
\begin{equation}
  T \chi = K \chi \ .
\label{eq:HBSeq}
\end{equation}
The kinetic part $T$ is given by 
\begin{equation}
  T(p;q) =  i S_F^{-1}(p + q/2) \otimes i S_F^{-1}(p - q/2) \ ,
\label{def T}
\end{equation}
where $S_F$ is the full fermion propagator 
$(\, i S_F^{-1}(p) = \fsl{p} - \Sigma \,)$, 
and the BS kernel $K$ in the improved ladder approximation 
is expressed as
\begin{equation}
  K(p;k) \ =\    \frac{N_c^2 - 1}{2 N_c} \ 
         \frac{\bar g^2(p,k)}{(p-k)^{2}}\ 
         \left( g_{\mu\nu} - \frac{(p-k)_\mu (p-k)_\nu}{(p-k)^2}
         \right) \cdot \gamma^\mu \otimes \gamma^\nu .
\end{equation} 
In the above expressions we used
the tensor product notation
\begin{equation}
  (A \otimes B) \,\chi  =  A\, \chi\, B \ ,
\end{equation}
and the inner product notation 
\begin{equation}     K \chi\ (p;q) =   
  \int \frac{d^4 k}{i(2\pi)^4}\  K(p,k)\  \chi(k;q) \ .
\end{equation}

It should
be noticed that the fermion propagators included in $T$ in
Eq.~(\ref{def T})
have complex-valued
arguments after the Wick rotation.
The arguments of the mass functions
appearing in two legs of the BS amplitude are expressed as
\begin{equation}
  - ( p \pm q/2 )^2 = u^2 + x^2 - \left( \frac{M_B}{2} \right)^2 
                      \mp i u M_B .
\end{equation}
In general, it is difficult to obtain mass
functions for
complex arguments. 
However, as we have shown 
in section~\ref{sec:complex_SD},
it is easy to obtain them
in the case of large $N_f$ QCD.

We now modify Eq.~(\ref{eq:HBSeq}) so that we can solve it 
numerically.
\footnote{In the following we explain the method for the vector and
axial-vector bound states.
The extension to the scalar and pseudoscalar bound states are easily
done.
} 
It is convenient to introduce
the conjugate bispinor bases defined by
\begin{equation}
  \bar\Gamma^i(p;q,\epsilon) \equiv
  \gamma_0 \Gamma^i(p^\ast;q,\epsilon)^\dag \gamma_0 \ .
\end{equation}
Multiplying 
these conjugate bispinor bases 
from the left, taking the trace 
of spinor indices, and summing over the polarizations, 
we rewrite
Eq.~(\ref{eq:HBSeq}) into the following form:
\begin{equation}
  T_{ij}(u,x) \chi^j(u,x) = \int \frac{y^2 \ dy \ dv}{8 \pi^3}
  K_{ij}(u,x;v,y) \chi^j(v,y) ,
\end{equation}
where 
the summation over the index $j$ is understood, and
\begin{eqnarray}
  T_{ij}(u,x) &=& \sum_{\epsilon} \frac{1}{3} 
        \mbox{tr} \left[ 
	\bar\Gamma^i(p;q,\epsilon) T(p;q) \Gamma^j(p;q,\epsilon)
	\right] ,\\
  K_{ij}(u,x;v,y) &=& \int_{-1}^{1} d\cos\theta \ \sum_{\epsilon} 
	\frac{1}{3} 
        \mbox{tr} \left[ \bar\Gamma^i(p;q,\epsilon)
	K(p,k) \Gamma^j(k;q,\epsilon) \right] \ ,
\end{eqnarray}
with the real variables $v$ and $y$ introduced as
$k\cdot q = i v M_B $ and $k\cdot p= - u v - x y \cos\theta$.
Here, we note that although the mass function $\Sigma(x)$ has 
the branch cut on the time-like axis as mentioned in 
section \ref{sec:complex_SD}, $T_{ij}$ has no singularity 
and becomes a continuous function for all the 
range of $u$ and $x$.
As for $K_{ij}$, the branch cut of running coupling $\bar{g}$ 
does not matter since its argument $(p_E^2 + k_E^2)$ never takes 
a negative value.

Using the property of $\chi^i$ in Eq.~(\ref{eq:even_chi}), we
restrict
the integration range as $v > 0$:
\begin{equation}
  \int dv K_{ij}(u,x;v,y) \chi^j(v,y) = 
	\int_{v > 0} dv \left[ K_{ij}(u,x;v,y) +
	 K_{ij}(u,x;-v,y) \right] \chi^j(v,y).
\end{equation}
Then,
all the variables $u$, $x$, $v$, and $y$
can be treated as positive values.

To discretize the
variables $u$, $x$, $v$, and $y$
we introduce new variables 
$U$, $X$, $V$, and $Y$ as
\begin{eqnarray}
  u = \Lambda\ e^{U/\Lambda}\ , & & x = \Lambda\ e^{X/\Lambda} \ , \nonumber\\
  v = \Lambda\ e^{V/\Lambda}\ , & & y = \Lambda\ e^{Y/\Lambda} \ ,
\end{eqnarray}
and
set UV and IR cutoffs as
\begin{equation}
  U, V \  \in \  [\lambda_U,\Lambda_U] ,\ \ 
  X, Y \  \in \  [\lambda_X,\Lambda_X] .
\end{equation}
We discretize the variables $U$ and $V$ 
into $N_{BS,U}$ points evenly, and $X$ and $Y$
into $N_{BS,X}$ points.
Then,
the original variables are labeled as
\begin{eqnarray}
  & & u_{[I_U]} = \Lambda 
                  \exp\left[\lambda_U/\Lambda + D_U I_U \right], \ \ 
      x_{[I_X]} = \Lambda 
                  \exp\left[\lambda_X/\Lambda + D_X I_X \right], \nonumber\\
  & & v_{[I_V]} = \Lambda 
                  \exp\left[\lambda_U/\Lambda + D_U I_V \right], \ \ 
      y_{[I_Y]} = \Lambda
                  \exp\left[\lambda_X/\Lambda + D_X I_Y \right], \nonumber
\end{eqnarray}
where
$I_U, I_V = 0, 1, 2, \cdots (N_{BS,U}-1)$ and
$I_X, I_Y = 0, 1, 2, \cdots (N_{BS,X}-1)$.
The measures $D_U$ and $D_X$ are defined as
\begin{equation}
  D_U = \frac{(\Lambda_U - \lambda_U)/\Lambda}{N_{BS,U} - 1}\ ,\ \ 
  D_X = \frac{(\Lambda_X - \lambda_X)/\Lambda}{N_{BS,X} - 1} \ .
\end{equation}
As a result, the integration is converted into the 
summation:
\begin{equation}
  \int_{v > 0} y^2 \ dy\ dv\ \cdots  \ \ \  \Longrightarrow \ \ \  
  D_U D_V \sum_{I_V,I_Y} v y^3 \ \cdots.
\end{equation}
In
order to avoid integrable singularities
in the kernel $K(u,x;v,y)$ at $(u,x)=(v,y)$, we adopt 
the following four-splitting prescription~\cite{Aoki:1990yp}:
\begin{eqnarray}
  K_{ij}(u,x,v,y) \ \  &\Longrightarrow& \ \   \frac{1}{4} 
      \ [\ K_{ij}(u,x,v_+,y_+) + K_{ij}(u,x,v_+,y_-) \nonumber\\
  & & \ \ \ \ +\  K_{ij}(u,x,v_-,y_+) + K_{ij}(u,x,v_-,y_-)\ ] ,
\end{eqnarray}
where    
\begin{equation}
  v_\pm = \exp\left[V \pm \frac{D_U}{4}\right] , \ \ 
  y_\pm = \exp\left[Y \pm \frac{D_X}{4}\right] . 
\end{equation}

\subsection{Fictitious eigenvalue method}
\label{ssec:FEM}

Now that all the variables have become discrete and 
the original integral equation (\ref{eq:HBSeq})
turned into a 
linear algebraic one, we are able to deal it numerically.
 However, 
it is difficult to find the bound state mass $M_B$ and the
corresponding BS amplitude $\chi$
directly 
since the HBS equation 
depends 
on
$M_B$ nonlinearly.
A way 
which enables us to solve the nonlinear eigenvalue 
problem is 
the {\it fictitious eigenvalue method}~\cite{Nakanishi:ph}.
In this method
we introduce a fictitious eigenvalue $\lambda$ 
and interpret the HBS equation~(\ref{eq:HBSeq})
as a linear eigenvalue
equation for a given bound state mass $M_B$:
\begin{equation}
  T \chi = \lambda \cdot K \chi.
\label{eq:fictitious HBS}
\end{equation}
Consequently, the HBS equation turns into 
an ordinary
eigenvalue problem 
which we 
can solve by standard algebraic techniques. 
By adjusting an input mass $M_B$ such that an eigenvalue $\lambda$
equals unity, we obtain the bound state mass and the corresponding BS
amplitude as a solution of the original HBS equation~(\ref{eq:HBSeq}).
In Appendix~\ref{app:positronium}, 
to show the validity of this method, 
we calculate the mass of the positronium using this method.

\section{Numerical Analysis}
\label{sec:Numerical_calculations}

In this section we show the results of our numerical analysis.

\subsection{Pseudoscalar bound state}
\label{sec:pseudoscalar}
As discussed
in subsection~\ref{sec:ladderSD}, the approximation 
to the argument of the running 
coupling in Eq.~(\ref{eq:approx_of_argument})
breaks the chiral symmetry 
explicitly \cite{Kugo:1992pr}. So, before 
solving the HBS equation for 
the massive bound states, we 
solve that for the pseudoscalar bound state and see how much
the chiral symmetry is explicitly broken by 
this approximation.

The mass of the lowest-lying pseudoscalar bound state should become 
zero because it appears as a Nambu-Goldstone boson 
when the chiral 
symmetry is spontaneously broken. So, we substitute zero for the 
bound state mass  
and check whether the fictitious eigenvalue 
$\lambda$ becomes unity.

We use the following parameters for the calculations:
\begin{equation}
  [\ \lambda_U ,\ \Lambda_U \ ]  
    =   [\ -18.0,\ 0\ ] ,\ \ 
  [\ \lambda_X ,\ \Lambda_X \ ]
    =   [\ -8.5,\ 0\ ] ,
\end{equation}
\begin{equation}
  N_{BS,U} \,=\, N_{BS,X} \,=\,  30 \ .
\end{equation}
We calculate
the fictitious eigenvalues for several values 
of $\alpha_\ast$ and show them
in Table~\ref{tab:pseudoscalar}.
\begin{table}
\begin{center}
\begin{tabular}{| c | c || c | c |}
\hline
   \ \ \ $\alpha_\ast$\ \ \  & \ \ \ \ \ $\lambda$\ \ \ \ \  
  &\ \ \ $\alpha_\ast$\ \ \  & \ \ \ \ \ $\lambda$\ \ \ \ \ \\
\hline\hline
  0.89  & 1.00121 & 0.95 & 1.00262 \\
\hline
  0.90 & 1.00205 & 0.96 & 1.00267 \\
\hline
  0.91 & 1.00230 & 0.97 & 1.00273 \\
\hline
  0.92 & 1.00241 & 0.98 & 1.00279 \\
\hline
  0.93 & 1.00249 & 0.99 & 1.00284 \\
\hline
  0.94 & 1.00255 & 1.00 & 1.00290 \\
\hline
\end{tabular}
\caption{
Fictitious eigenvalues 
obtained by solving Eq.~(\ref{eq:fictitious HBS}) 
for the pseudoscalar bound state
with zero mass used as an input. 
}
\label{tab:pseudoscalar}
\end{center}
\end{table}
We can see that
$\lambda = 1$ is satisfied within $0.3 \%$ uncertainty.  
This implies that our calculations actually reproduce the massless
Nambu-Goldstone boson within the numerical error, and that the
effect of explicit chiral symmetry breaking caused by the
approximation for the running coupling is negligible.

\subsection{Vector, axial-vector, and scalar bound states}
\label{ssec:VB}

In this subsection we show the results of the numerical 
calculations for the masses of the vector, axial-vector, 
and scalar  bound states.
For the UV and IR cutoffs 
we adjust the values of them in such a way that
the dominant supports of the integrands of the decay constant
in Eq.~(\ref{eq:F_V}) and the normalization condition
in Eq.~(\ref{eq:normalization})
lie in the energy region between the UV and IR cutoffs.
As an example, we show the integrands of the decay constant 
and the normalization condition 
for the vector bound states in Appendix~\ref{app:Uncertainties}.
{}From these figures, 
the dominant supports lie in the lower energy region
for smaller value of $\alpha_\ast$.
Then, we use the following 
$\alpha_\ast$-dependent UV and IR cutoffs for 
the vector and the axial-vector bound states:
\begin{eqnarray}
  \left[\ \lambda_U ,\ \Lambda_U \ \right]  
    &=&   [\ -12.0 + 22.0 \times (\alpha_\ast - 1.0) 
        \ ,\ -1.0 + 35.0 \times (\alpha_\ast - 1.0 ) \ ] , 
\label{eq:IR cutoffs}
\\
  \left[\ \lambda_X ,\ \Lambda_X \ \right]
    &=&   [\ -5.0 + 22.0 \times (\alpha_\ast - 1.0 )
        \ ,\ -2.0 + 20.0 \times (\alpha_\ast - 1.0 ) \ ] .
\label{eq:UV cutoffs}
\end{eqnarray}
For the scalar bound state, 
on the other hand, 
we use the following fixed UV and IR cutoffs:
\begin{equation}
  [\ \lambda_U ,\ \Lambda_U \ ]  
    =   [\ -18.0,\ 0\ ] ,\ \ 
  [\ \lambda_X ,\ \Lambda_X \ ]
    =   [\ -10.0,\ 0\ ] .
\end{equation}
Although the integrands of the normalization conditions are 
shown in Appendix~\ref{app:Uncertainties} only for the vector 
bound states, we have checked that the dominant supports 
always lie within the energy region between UV and IR cutoffs 
for all kinds of bound states and for all values of $\alpha_\ast$.
As for the numbers of the discretization,
we use 
\begin{equation}
N_{BS,U} = 20 \ , \quad N_{BS,X} = 55 ,
\label{NBS num}
\end{equation}
for the vector and the axial-vector bound states, and 
\begin{equation}
  N_{BS,U} \,=\, N_{BS,X} \,=\,  30 ,
\end{equation}
for the scalar bound states.
In Appendix~\ref{app:Uncertainties}, 
we show that these numbers of discretization are large 
enough for the present analysis.

We should stress that we actually found a solution for  
Eq.~(\ref{eq:fictitious HBS}) reproducing $\ \lambda = 1$ 
for all the types of the bound states  
in the range of $\alpha_\ast \in [0.885:1]$.
This means that there do exist the vector, axial-vector, 
and scalar
bound states near the phase transition point in the broken phase.
\footnote{
On the other hand, we cannot find any solutions for the HBS equations 
in the symmetric phase, i.e., $\alpha_\ast < \alpha_{\rm cr}$ 
(or, equivalently,  $N_f >N_f^{\rm crit}$). 
This fact seems consistent with the property of 
the conformal phase transition which has no bound states 
in the symmetric phase~\cite{Miransky:1996pd,Appelquist:1998rb}.
}

\begin{figure}
  \begin{center}
    \includegraphics[height=8cm]{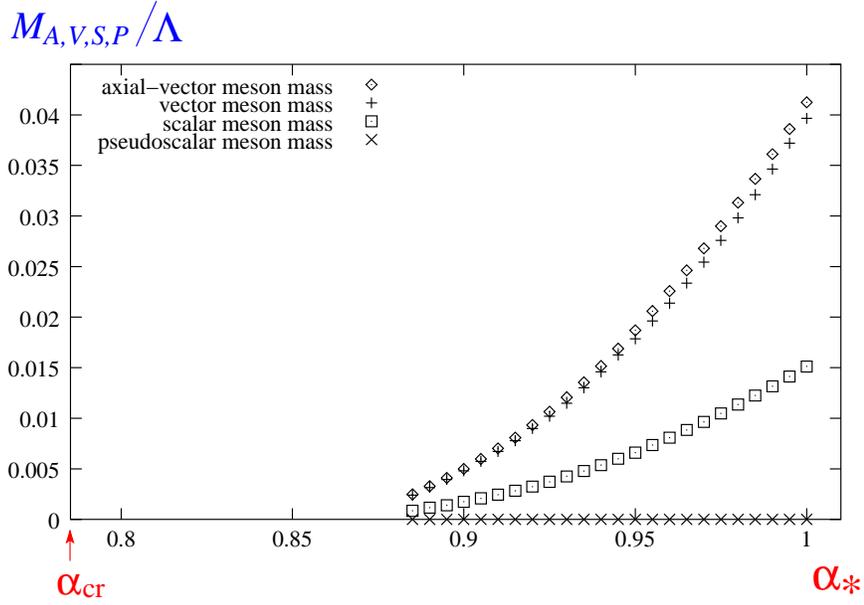}
  \end{center}
\caption[]{
  Values of the scalar, pseudoscalar, 
  vector, and axial-vector 
  meson masses 
  for several values of 
  $\alpha_\ast$.
  }
\label{fig:all_masses}
\end{figure}

\begin{figure}
  \begin{center}
    \includegraphics[height=8cm]{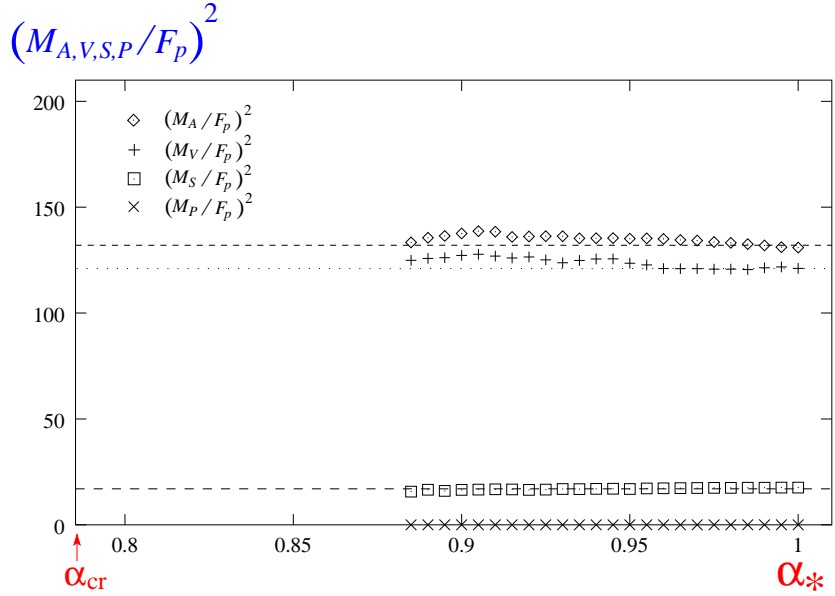}
  \end{center}
\caption[]{
  Values
  of $(M_A / F_P)^2 $, $(M_V / F_P)^2 $, $(M_S / F_P)^2 $ and 
  $(M_P / F_P)^2$  for several values
  of $\alpha_\ast$ (indicated by $\Diamond$, $+$, $\Box$, and
  $\times$, 
  respectively). 
  Dotted lines represent the values of $(M_A / F_P)^2 = 17 $, 
  $(M_V /  F_P)^2 = 121$, and $(M_S / F_P)^2 = 132$ 
  obtained from Eqs.~(\ref{eq:d_Mv}) 
  and (\ref{eq:d_fpi}).
  }
\label{fig:M_over_Fpi}
\end{figure}

Now, let us show the critical behavior of the masses of the existing 
bound states.
In Fig.~\ref{fig:all_masses},
we plot all the bound state masses calculated 
for several values of $\alpha_\ast$ together with the pseudoscalar 
meson masses obtained in the previous subsection.
This figure shows that the masses of the vector, axial-vector, 
and scalar
bound states go to zero simultaneously 
as the coupling approaches its critical value (or, equivalently,
$N_f \rightarrow N_f^{\rm crit}$) :
\begin{equation}
M_S, M_V, M_A \ \rightarrow\  0 \  \quad \mbox{for} \ \ \alpha_\ast
\ \rightarrow\ \alpha_{\rm cr} \ .
\end{equation}
Next, to study the critical behavior of $M_S$, $M_V$, and $M_A$ 
we use the function of
the form in Eq.~(\ref{eq:fitting}) and fit the value of $d$ by
minimizing
\begin{equation}
\sum_{\alpha_\ast} 
\left\vert M_{S,V,A}(\alpha_\ast) - h(\alpha_\ast) \right\vert^2
\ .
\end{equation}
The resultant best fitted values of $d$ for the scalar, vector, and 
axial-vector bound states are 
\begin{equation}
  d_{M_S} \simeq 6.2 \ ,\ \   d_{M_V} \simeq 16.5 
  \ ,\ \  d_{M_A} \simeq 17.2\  ,
\label{eq:d_Mv}
\end{equation}
respectively.
We also plot (the square of) the ratio of the bound state 
masses to $F_P$ for several values of $\alpha_\ast$ in
Fig.~\ref{fig:M_over_Fpi}.
The dotted lines plotted together with the data in this figure 
represent the values of the following ratios obtained 
from Eqs.~(\ref{eq:d_Mv}) and (\ref{eq:d_fpi}) :
\begin{equation}
 \left( M_S / F_P \right)^2 =  17 \ ,\ \ 
 \left( M_V / F_P \right)^2 = 121 \ ,\ \ 
 \left( M_A / F_P \right)^2 = 132 \ .
\label{eq:M_over_F-square}
\end{equation}
This figure clearly shows that all the masses of 
the scalar, vector, and axial-vector bound states have the same 
scaling property as that of $F_P$ :
\begin{equation}
  \frac{M_{S,V,A}}{F_P}\  \sim \ {\rm constant}.
\end{equation}
We can also say that these masses have the same scaling property as 
that of $\Sigma(m^2)$ since $F_P$ and $\Sigma(m^2)$ have the 
same scaling property.
Their ratios are summarized as follows:
\begin{equation}
  \Sigma^2(m^2) :  M_S^2 :  M_V^2  :  M_A^2\ \  =\ \  1\  :\  2.4\  :\
  17.0\  :\  18.5\ .
\end{equation}
One might think that the vector and axial-vector boundstates
decay into a fermion and an antifermion since 
$\ M_V^2 > 4 \Sigma^2(m^2)\ $ and $\ M_A^2 > 4 \Sigma^2(m^2)$.
However, this does not happen: As we noticed above 
Eq.~(\ref{eq:Sigma_m}), 
$\ \Sigma(m^2)\ $ is not the pole mass but the dynamical mass 
defined in the space-like region.  Furthermore, as we have
shown in section~\ref{sec:complex_SD}, 
the fermion propagators do not have any
poles in the entire complex plane including the time-like
axis where the pole mass of fermion should be defined.  
Thus the vector and axial-vector boundstates do not decay into
a fermion and an antifermion.

\section{Conclusion and Discussion}
\label{sec:Discussion_and_conclusion}

In this paper we first pointed out that, 
when we solve the Schwinger-Dyson (SD) equation in large $N_f$ QCD, 
we do not need to introduce any infrared regularizations for the
running coupling since it takes a finite value for 
all the range of the energy region due to the existence of the IR
fixed point.  
In the case of small $N_f$, we have to regularize the infrared
divergence of the running coupling, and different schemes of 
regularizations would give different results.
Furthermore,
it is difficult to find the regularization which makes the
analytic structure of the running coupling simple enough.
On the contrary, 
the solution of the two-loop RGE in large $N_f$ QCD is explicitly 
written in terms of the Lambert $W$ function, 
and the running coupling does not have any singularities on the
complex plane except for the time-like axis
when $N_f$ is close to $N_f^{\rm crit}$.
This significant feature of the running coupling in large $N_f$ QCD 
enabled us to solve the SD equation on the complex plane. 

Then, we solved the homogeneous Bethe-Salpeter (HBS) equations
for the scalar, pseudoscalar, vector, and axial-vector bound states 
of quark and anti-quark in large $N_f$ QCD 
with the improved ladder approximation in the Landau gauge.
In the quark propagator included in the HBS equation,
we used the
quark mass function obtained from the SD equation
with the same approximation, which is needed for the consistency with
the chiral symmetry.

We first showed that the HBS equation provides
the massless pseudoscalar bound state in the broken phase
which is identified with the Nambu-Goldstone boson 
associated with the spontaneous breaking of the chiral symmetry.
Next, we 
showed that 
there actually exist 
vector, axial-vector, and scalar bound states 
even near the phase transition point in the broken phase,
and that
their masses 
decreases as the number of massless 
quarks $N_f$ increases.
At the critical point all the masses go to zero, 
showing the 
same scaling property 
as that of the pseudoscalar meson decay constant $F_P$
consistently with the picture expected 
from the conformal phase 
transition~\cite{Miransky:1996pd,Chivukula:1996kg}.

Let us discuss 
the pattern of the chiral symmetry 
restoration
by considering the representation of chiral 
$\mbox{SU}(N_f)_{\rm L}\times\mbox{SU}(N_f)_{\rm R}$
of the low-lying 
mesons extending the analyses done in 
Refs.~\cite{Gilman:1967qs,Weinberg:hw}.

For $N_f=3$ 
the pseudoscalar meson denoted by $\pi$
and the longitudinal axial-vector meson denoted by $A_1$
are an admixture of $(8\,,\,1) \oplus(1\,,\,8)$ and 
$(3\,,\,3^*)\oplus(3^*\,,\,3)$, since the 
chiral symmetry is spontaneously
broken~\cite{Gilman:1967qs,Weinberg:hw}
\begin{eqnarray}
\vert \pi\rangle &=&
\vert (3\,,\,3^*)\oplus (3^*\,,\,3) \rangle \sin\psi
+
\vert(8\,,\,1)\oplus (1\,,\,8)\rangle  \cos\psi
\ ,
\nonumber
\\
\vert A_1(\lambda=0)\rangle &=&
\vert (3\,,\,3^*)\oplus (3^*\,,\,3) \rangle \cos\psi 
- \vert(8\,,\,1)\oplus (1\,,\,8)\rangle  \sin\psi
\ ,
\label{mix pi A}
\end{eqnarray}
where $\lambda$ denotes the helicity in the collinear 
frame, and the experimental value of the mixing angle $\psi$ is 
given by approximately 
$\psi=\pi/4$~\cite{Weinberg:hw}.
On the other hand, the longitudinal vector meson denoted by $\rho$
belongs to pure $(8\,,\,1)\oplus (1\,,\,8)$
and the scalar meson denoted by $S$ to 
pure $(3\,,\,3^*)\oplus (3^*\,,\,3)$:
\begin{eqnarray}
\vert \rho(\lambda=0)\rangle &=&
\vert(8\,,\,1)\oplus (1\,,\,8)\rangle  
\ ,
\nonumber
\\
\vert S\rangle &=&
\vert (3\,,\,3^*)\oplus (3^*\,,\,3) \rangle 
\ .
\label{rhoS}
\end{eqnarray}

When the chiral symmetry is restored at the
phase transition point, it is natural 
that the 
chiral representations coincide with the mass eigenstates:
The representation mixing is dissolved.
{}From Eq.~(\ref{mix pi A}) one can easily see
that
there are two ways to express the representations in the
Wigner phase of the chiral symmetry:
The conventional manifestation \'{a} la the linear sigma model
(called the GL manifestation in Ref.~\cite{HY:PRep})
corresponds to 
the limit $\psi \rightarrow \pi/2$ in which
$\pi$ is in the representation
of pure $\ (N_f\,,\,N_f^\ast)\oplus(N_f^\ast\,,\,N_f)\ $ of
$\ \mbox{SU}(N_f)_{\rm L} \times\mbox{SU}(N_f)_{\rm R}\ $
together with the scalar meson, 
both being the chiral partners:
\begin{eqnarray}
\mbox{(GL)}
\qquad
\left\{
\begin{array}{rcl}
\vert \pi\rangle\,, \vert S\rangle
 &\rightarrow& 
\vert  (N_f\,,\,N_f^\ast)\oplus(N_f^\ast\,,\,N_f)\rangle\ ,
\\
\vert A_1(\lambda=0) \rangle \,,
\vert \rho (\lambda=0) \rangle  &\rightarrow&
\vert(N_f^2-1\,,\,1) \oplus (1\,,\,N_f^2-1)\rangle\ .
\end{array}\right.
\end{eqnarray}
On the other hand, the vector manifestation 
(VM)~\cite{Harada:2000kb} corresponds 
to the limit $\psi\rightarrow 0$ in which the $A_1$ 
goes to a pure 
$(N_f\,,\,N_f^\ast)\oplus(N_f^\ast\,,\,N_f)$, now degenerate with
the scalar meson in the same representation, 
but not with $\rho$ in 
$(N_f^2-1\,,\,1) \oplus (1\,,\,N_f^2-1)$:
\begin{eqnarray}
\mbox{(VM)}
\qquad
\left\{
\begin{array}{rcl}
\vert \pi\rangle\,, \vert \rho (\lambda=0) \rangle
 &\rightarrow& 
\vert(N_f^2-1\,,\,1) \oplus (1\,,\,N_f^2-1)\rangle\ ,
\\
\vert A_1(\lambda=0)\rangle\,, \vert S\rangle  &\rightarrow&
\vert  (N_f\,,\,N_f^\ast)\oplus(N_f^\ast\,,\,N_f)\rangle\ .
\end{array}\right.
\end{eqnarray}
Namely, the
degenerate massless $\pi$ and (longitudinal) $\rho$ at the 
phase transition point are
the chiral partners in the
representation of $(N_f^2-1\,,\,1) \oplus (1\,,\,N_f^2-1)$.

Now, what does our result say on the chiral representation of
low-lying mesons?
As can be seen from Fig.~\ref{fig:M_over_Fpi},
the resultant values of the masses
obtained 
from the HBS equation roughly satisfies the following 
relation~\cite{Weinberg:hw}:
\begin{equation}
M_A^2 + M_P^2 = M_V^2 + M_S^2,
\label{sat cond}
\end{equation}
for all values of $\alpha_\ast$.
This relation holds independently of the mixing angle $\psi$
given in Eq.~(\ref{mix pi A}) when the low-lying mesons saturate the
chiral algebra shown in Ref.~\cite{Weinberg:hw}.
Then, 
it is reasonable to discuss the chiral representation 
without worrying about the influence of the excited states of
the bound states. 
By using the relation $\tan{\psi} = M_V/M_S$~\cite{Weinberg:hw}
and 
the values of $M_V$ and $M_S$
obtained from the HBS equation in 
subsection~\ref{ssec:VB},
the value of the mixing angle $\psi$ is roughly determined as
\begin{equation}
  \ \tan{\psi} = M_V/M_S \sim 3 .
\end{equation}
This implies that 
$\pi$ and the longitudinal $A_1$ are still admixtures of the
pure chiral representation even at the chiral restoration point:
\begin{eqnarray}
  \vert \pi \rangle &\rightarrow& 
        \vert (N_f , N_f^\ast) \oplus (N_f^\ast , N_f)\  \rangle 
        \sin{\psi}  
  \ +\  \vert (N_f^2 - 1 , 1)  \oplus (1 , N_f^2 - 1) \  \rangle 
        \cos{\psi},
\nonumber
\\
  \vert A_1 \rangle &\rightarrow& 
        \vert (N_f , N_f^\ast) \oplus (N_f^\ast , N_f)\  \rangle 
        \cos{\psi}  
  \ -\  \vert (N_f^2 - 1 , 1)  \oplus (1 , N_f^2 - 1) \  \rangle 
        \sin{\psi}.
\nonumber\\
\label{new pattern}
\end{eqnarray}
This may suggest the existence of a new type of manifestation of
chiral symmetry restoration in large $N_f$ QCD
which is neither of the GL manifestation nor the 
simple version of the vector manifestation (VM).

Several comments are in order 

In Appendix~\ref{app:decay_const}, we show the calculations of 
the coupling constants $F_V$, $F_A$, and 
$G_S$ of the vector, axial-vector, and scalar bound states 
to the vector current, axial-vector current, and scalar density.
The results shows that they also have the same scaling properties 
as $F_P$.
These results indicate that all the dimension-full quantities 
determined by the infrared dynamics have the same scaling properties,
as far as the (improved) ladder approximation is concerned.

Although the masses obtained from the HBS equation satisfy the
condition (\ref{sat cond}) needed for the saturation of the
chiral algebra, the couplings $F_P$, $F_V$, and $F_A$ do not
seem to 
satisfy the first Weinberg's sum rule~\cite{Weinberg:SR}:
$F_P^2 + F_A^2 = F_V^2$.
We have not fully understood what this means for the
pattern of chiral symmetry restoration.
Apparently, reducing the numerical uncertainty will help us to
reach the final understanding.

In the present analysis we did not include the effect from
the four-fermion interaction which is induced in the 
case of $\gamma_m \simeq 1$ as was conjectured
in strong coupling QED~\cite{Leung-Love-Bardeen}
and was demonstrated in walking gauge theories~\cite{KSY91,Aoki}.
It is not clear at this moment whether or not 
the qualitative results in the present
analysis will 
be changed 
when we include such an effect.

In the present analysis we stressed that the running coupling
in large $N_f$ QCD determined from a two-loop $\beta$-function
is expressed as the Lambert $W$ function which enables us to 
solve the HBS and SD equations with mutual consistency near
the critical point.
Apparently, this Lambert $W$ function can be used
as an
infrared regularization to solve the HBS and SD equations
with mutual consistency in the case of QCD with small $N_f$.
It will be very interesting to 
study meson masses near
the chiral phase transition in hot and/or dense QCD
by using such an infrared regularization.

We think that it is important
to clarify which effective field theory (EFT)
describes the new pattern
of the chiral symmetry restoration expressed in 
Eq.~(\ref{new pattern}).
Especially, it is very interesting to see how the matching
between the EFT and the underlying QCD with large $N_f$ can be
done to determine bare parameters of the Lagrangian of the EFT.

\section*{Acknowledgments}

This work was supported in part by the JSPS Grant-in-Aid for the
Scientific Research (B)(2) 14340072.
The work of M.H. was supported in part by the
Brain Pool program (\#012-1-44) provided by the Korean Federation
of Science and Technology Societies and USDOE Grant
\#DE-FG02-88ER40388.  
M.H. would like to thank Professor Mannque Rho, 
Professor Gerry Brown, and Professor Dong-Pil Min for
their hospitality during his stays at KIAS, 
at SUNY at Stony Brook, and at Seoul National University 
where part of this work was done.


\appendix

\renewcommand\theequation{\Alph{section}.\arabic{equation}}

\begin{flushleft}
\LARGE\bf Appendices
\end{flushleft}

\section{Positronium} 
\label{app:positronium}

In this appendix, to show the validity of 
the fictitious eigenvalue method
for solving the HBS equation
explained in subsection~\ref{ssec:FEM}, 
we
calculate the mass of the ortho-positronium which is 
the vector bound state of the electron and the positron.
The same analysis was done in Ref.~\cite{Harada:1995nx}, and 
here we follow the analysis.

In the weak coupling limit the HBS equation for the 
ortho-positronium can be solved
analytically, 
and the energy spectrum takes the following form~\cite{Itzykson}:
\begin{equation}
    M_V^{(n)} \ =\   2\, m_e\  -\  
                   \frac{m_e \,\alpha^2}{4\,n^2},
\label{eq:positronium_mass}
\end{equation}
where $m_e$ is the mass of the electron and the positron,
and $\alpha=1/137$ is the coupling constant of QED.

\begin{figure}
  \begin{center}
    \includegraphics[height=6cm]{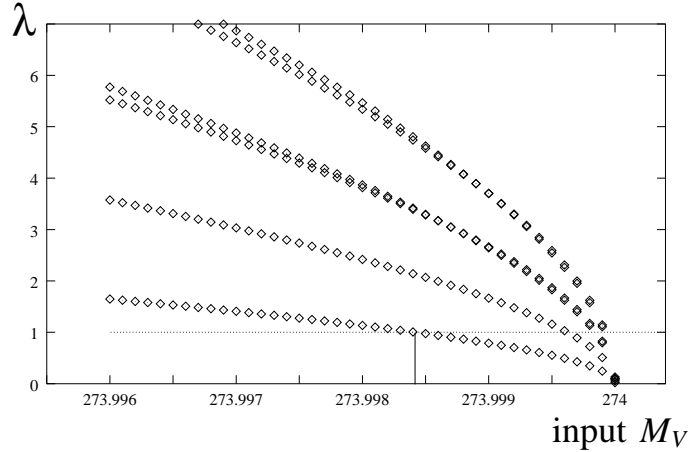}
  \end{center}
\caption{Fictitious eigenvalues for the positronium.}
\label{fig:positronium_BSsv}
\end{figure}

We use following parameters in our calculation:
\begin{equation}
  [\ \lambda_{U} ,\ \Lambda_{U} \ ]  
         =   [\ -18.5,\ -2.9\ ] \ ,\ \ 
  [\ \lambda_{X} ,\ \Lambda_{X} \ ]
         =   [\ -10.8,\ 2.2\ ] \ ,
\end{equation}
\begin{equation}
    N_{BS,U} \,=\, N_{BS,X} \,=\,  28 \ ,\ \ 
  m_e = 137.0 \ .
\end{equation}
(We used the energy scale which satisfies the relation 
$m_e \alpha = 1$ following Ref.~\cite{Harada:1995nx}.)
In 
Fig.~\ref{fig:positronium_BSsv},
we show the resultant values
of the
fictitious eigenvalue $\lambda$ 
for several values of
the input parameter 
$M_V$.
Finding
the point where the smallest $\lambda$ becomes unity, we determine
the value of the mass of the ground state
$M_V^{(1)}$
as the solution of the original HBS equation:
\begin{equation}
  M_V^{(1)}\ =\ 273.99842 .
\end{equation}
 {}From this the binding energy is calculated as
 \begin{equation}
  E^{(1)} \ = \ 0.00158 .
 \end{equation}
These values are in good agreement with the values
$\ M_V^{(1)} = 273.99817\ $ and $E^{(1)} = 0.00183$ 
derived from Eq.~(\ref{eq:positronium_mass}).
This shows that our numerical method
works well to obtain the mass of the ground state.

\section{Bispinor bases for scalar, pseudoscalar, vector, and 
axial-vector bound states}
\label{app:bispinor-bases}
In this appendix we show the explicit forms of the bispinor bases 
for the scalar, pseudoscalar, vector, and axial-vector bound states.
Here, we use the notation $\hat{q}_\mu = q_\mu / M_B $ 
with $M_B$ being the mass of the bound states, 
and $[a,b,c] \equiv a[b,c] + b[c,a] + c[a,b]$.

Bispinor base for the scalar bound state ($J^{PC} = 0^{++}$) 
is given by
\begin{eqnarray}
  \Gamma_{(S)}^1 =  {\bf 1} ,\ \ 
  \Gamma_{(S)}^2 =  \fsl{p} ,\ \ 
  \Gamma_{(S)}^3 =  \fsl{\hat{q}} (p \cdot \hat{q}) , \ \ 
  \Gamma_{(S)}^4 =  \frac{1}{2}\ [\fsl{p},\fsl{\hat{q}}] ,
\end{eqnarray}
and that for the pseudoscalar bound state ($J^{PC} = 0^{-+}$) 
is given by 
\begin{eqnarray}
  \Gamma_{(P)}^1 = \gamma_5 ,\ \ 
  \Gamma_{(P)}^2 = \fsl{p} \ (p \cdot \hat{q})\ \gamma_5 ,\ \ 
  \Gamma_{(P)}^3 = \fsl{\hat{q}} \ \gamma_5 , \ \ 
  \Gamma_{(P)}^4 = \frac{1}{2}\ [\fsl{p},\fsl{\hat{q}}]\ 
                   \gamma_5 \ .
\end{eqnarray}
Furthermore, for the vector bound state ($J^{PC} = 1^{--}$) we use
\begin{eqnarray}
& &  \Gamma_{(V)}^1 = \fsl{\epsilon} ,\ \ 
     \Gamma_{(V)}^2 = \frac{1}{2} [\fsl{\epsilon},\fsl{p}] 
                           (p \cdot \hat q) ,\ \ 
     \Gamma_{(V)}^3 = \frac{1}{2} [\fsl{\epsilon},\fsl{\hat q}] ,\ \ 
     \Gamma_{(V)}^4 = \frac{1}{3!}[\fsl{\epsilon},\fsl{p},\fsl{\hat
     q}] , 
     \\  
& &  \Gamma_{(V)}^5 = (\epsilon \cdot p) ,\ \ 
     \Gamma_{(V)}^6 = \fsl{p} (\epsilon \cdot p) ,\ \ 
     \Gamma_{(V)}^7 = \fsl{\hat q}(p \cdot \hat q) (\epsilon \cdot p) ,\ \ 
     \Gamma_{(V)}^8 = \frac{1}{2} [\fsl{p},\fsl{\hat q}](\epsilon \cdot p) ,
\nonumber
\label{V bases}
\end{eqnarray}
and for the axial-vector bound state
($J^{PC} = 1^{++}$) 
\begin{eqnarray}
  \Gamma_{(A)}^1 &=& \fsl{\epsilon}\ \gamma_5 ,\ \ \ 
     \Gamma_{(A)}^2 = \frac{1}{2} [\fsl{\epsilon},\fsl{p}] 
                           \gamma_5  ,\ \ \ 
     \Gamma_{(A)}^3 = \frac{1}{2} [\fsl{\epsilon},\fsl{\hat q}]\ (p \cdot \hat q)
                \ \gamma_5  ,\nonumber\\
     \Gamma_{(A)}^4 &=& \frac{1}{3!}[\fsl{\epsilon},\fsl{p},\fsl{\hat q}]
     \ \gamma_5 ,\ \ \ 
     \Gamma_{(A)}^5 = (\epsilon \cdot p)\ (p \cdot \hat q)\ \gamma_5  ,\ \ \ 
     \Gamma_{(A)}^6 = \fsl{p} (\epsilon \cdot p)\ \gamma_5  ,
     \nonumber\\
     \Gamma_{(A)}^7 &=& \fsl{\hat q}\ (\epsilon \cdot p)\ (p \cdot \hat q)
                \ \gamma_5  ,\ \ \ 
     \Gamma_{(A)}^8 = \frac{1}{2} [\fsl{p},\fsl{\hat q}](\epsilon \cdot p)\
     (p \cdot \hat q)\ \gamma_5  .
\end{eqnarray}

\section{Coupling constants to currents and scalar density}
\label{app:decay_const}

In this subsection we calculate coupling constants $F_V$, $F_A$, and 
$G_S$ of the vector, axial-vector, and scalar bound states 
to the vector current, axial-vector current, and scalar density.
They are defined by
\begin{eqnarray}
  \langle \ 0\  \vert \ \bar\psi(0)\ \gamma^\mu \ \frac{\lambda_a}{2}
             \ \psi(0) \ \vert \ V_b(q,\epsilon)
  \ \rangle \ &=&\ 
  \delta_{a b}\  F_V \ M_V \ \epsilon^\mu ,
\label{eq:f_V_def}\\
  \langle \ 0\  \vert \ \bar\psi(0)\ \gamma^\mu \ 
          \gamma_5\ \frac{\lambda_a}{2}
             \ \psi(0) \ \vert \ A_b(q,\epsilon)
  \ \rangle \ &=&\ 
  \delta_{a b}\  F_A \ M_A \ \epsilon^\mu ,\\
  \langle \ 0\  \vert \ \bar\psi(0)\ \frac{\lambda_a}{2}
             \ \psi(0) \ \vert \ S_b(q,\epsilon)
  \ \rangle \ &=&\ 
  \delta_{a b}\  G_S,
\end{eqnarray}
where $\lambda_a$ is the flavor matrix normalized as
$\mbox{tr} [ \lambda_a \lambda_b ]
= 2\delta_{a b}$.

\begin{figure}
  \begin{center}
    \includegraphics[height=5cm]{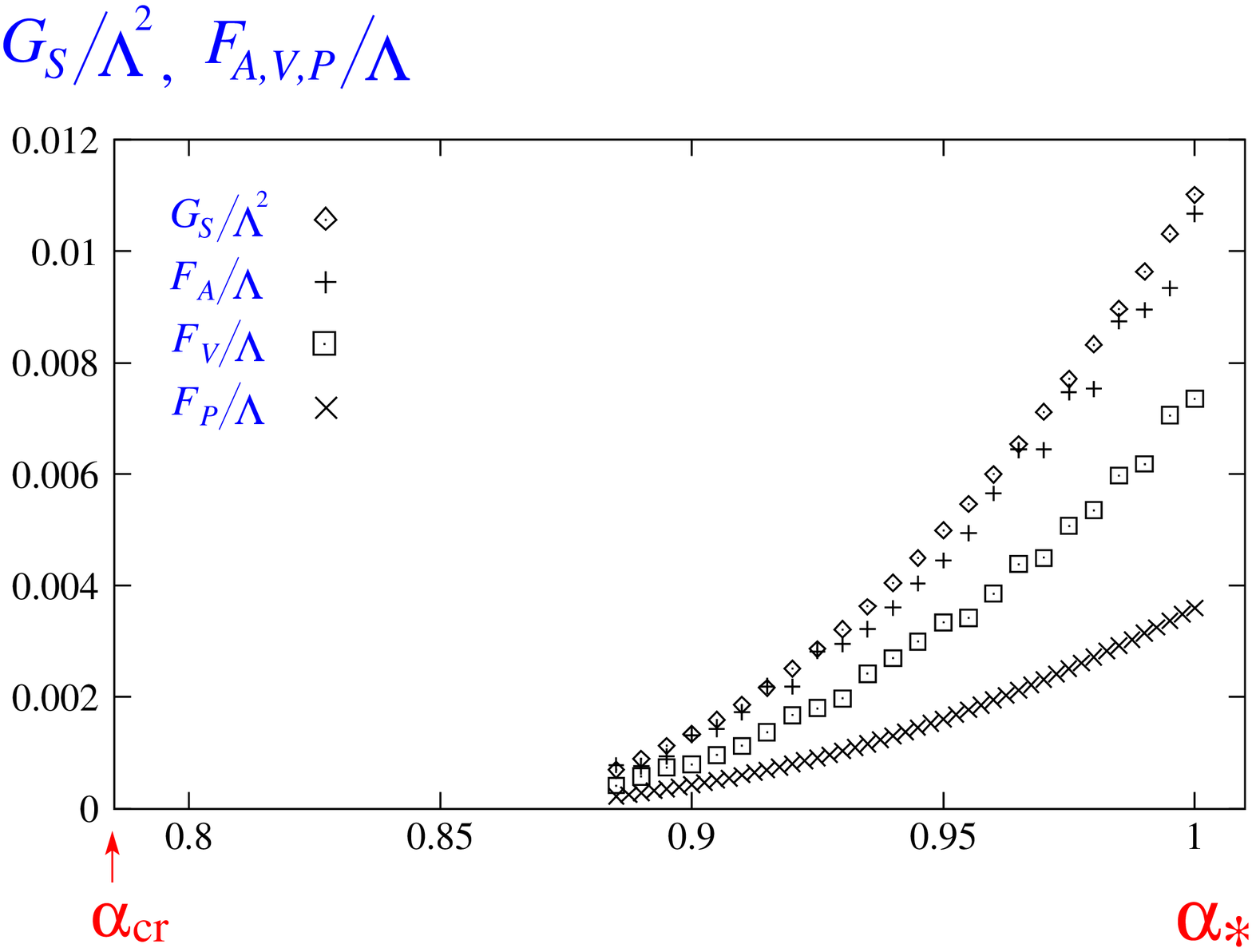}\hspace{1cm}
    \includegraphics[height=5cm]{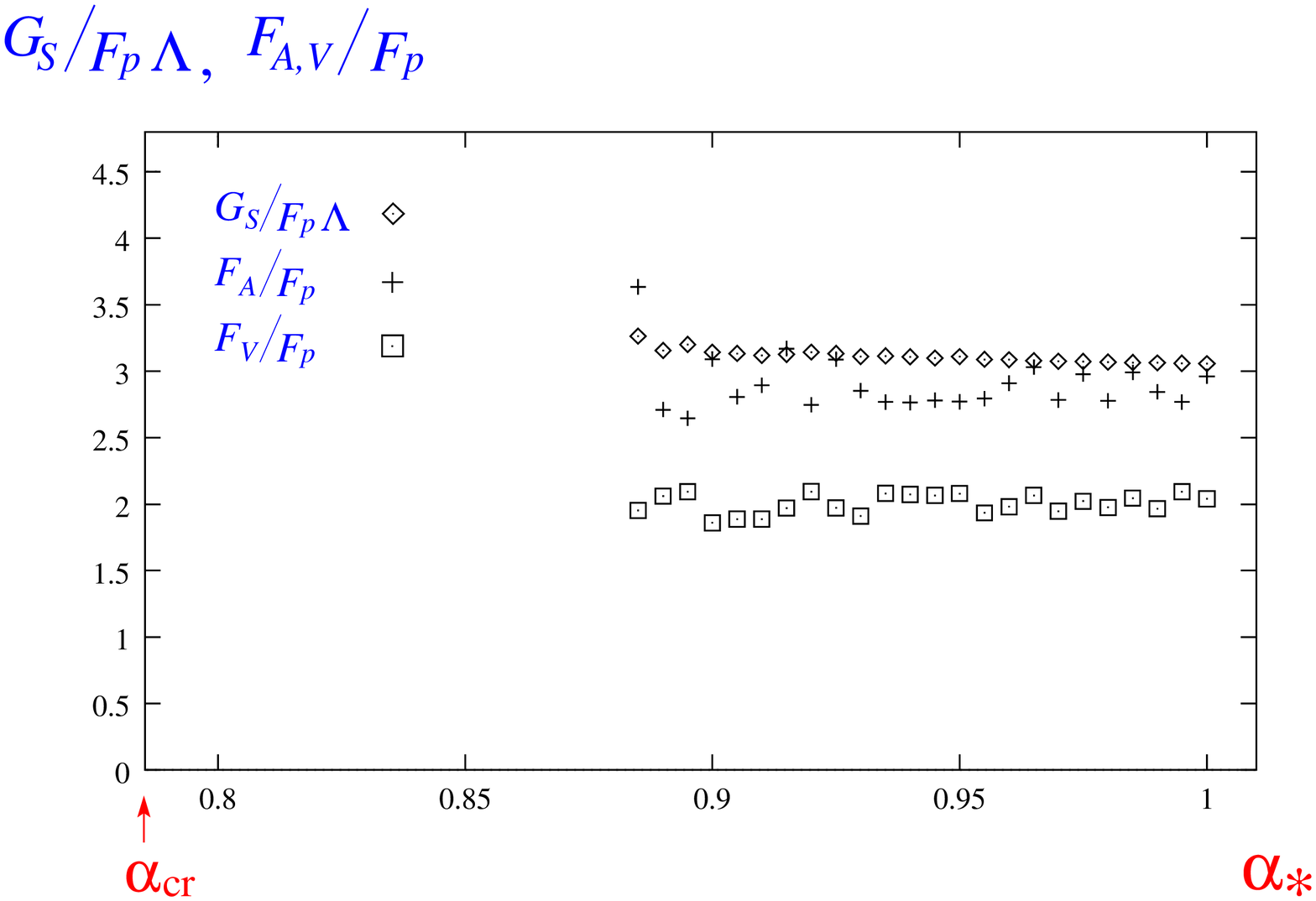}
    \ \\
    \hspace*{0cm} \ (a) \ \hspace{6.9cm} \ (b) 
  \end{center}
\caption{Values of \ (a)\ $F_V$, $F_A$, $G_S$, and $F_P$ 
and \ (b)\ $G_S / F_P$, $F_V / F_P$, and $F_A / F_P$ 
for several values of $\alpha_\ast$.
}
\label{fig:decay_const}
\end{figure}

By using the BS amplitudes for the vector bound state,
$F_V$ is expressed as
\begin{equation}
  F_V M_V \ =\ - \ \frac{\sqrt{2}\,i\,N_c }{\pi^3}\ 
    \int_{0}^{\infty}du 
    \int_{0}^{\infty}dx 
    \Bigg[\  x^2\,\chi^1(u,x) \ 
    -\ \frac{x^4}{3}\,\chi^6(u,x)
    \ \Bigg] \ .
\label{eq:F_V}
\end{equation}
In the above expression, the normalization of the BS amplitudes
$\chi^i$ are determined by the following normalization 
condition~\cite{Nakanishi:ph}:
\begin{eqnarray}
     2 M_V \,\delta_{\epsilon \epsilon '} = 
      i N_c
     \int \frac{d^4p}{(2\pi)^4} 
     \Bigg[\, \bar\chi(p;q,\epsilon)\,
     \frac{\partial T(p;q)}{\partial M_V}\,
     \chi(p;q,\epsilon ')\,\Bigg] \ .
\label{eq:normalization}
\end{eqnarray}
Here, we notice again that $T(p;q)$ has no singularity 
although the fermion propagator $S_F$ has a branch cut 
in the time-like region.
So the integral in Eq.~(\ref{eq:normalization}) is 
well-defined.
Once we have obtained $M_V$ and the corresponding BS amplitudes 
by solving the HBS equation, 
we can calculate $F_V$ from Eq.~(\ref{eq:F_V}).
We can also calculate $F_A$ and $G_S$ in a similar way.
In Fig.~\ref{fig:decay_const} (a) we show the values of $F_V$, $F_A$, 
and $G_S$ together with $F_P$ for several values of $\alpha_\ast$.
To see the scaling properties, we plot the ratio of 
$F_V$, $F_A$, and $G_S$ to $F_P$ in Fig.~\ref{fig:decay_const} (b).
This figure shows that $F_V$, $F_A$, and $G_S$ have 
the same scaling properties as that of $F_P$.

\section{Uncertainties for numerical calculations}
\label{app:Uncertainties}

\begin{figure}
  \begin{center}
    \includegraphics[width=6cm]{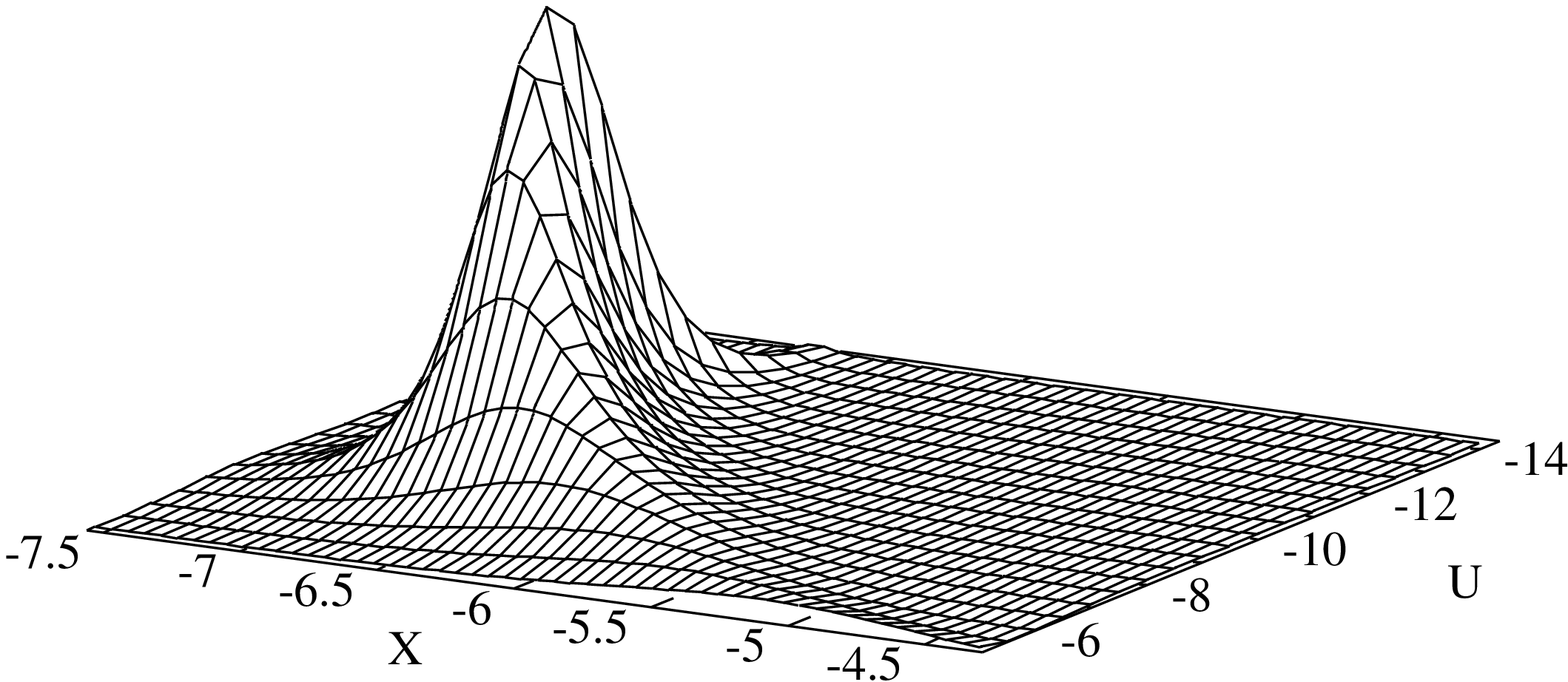}
    \ \includegraphics[width=6cm]{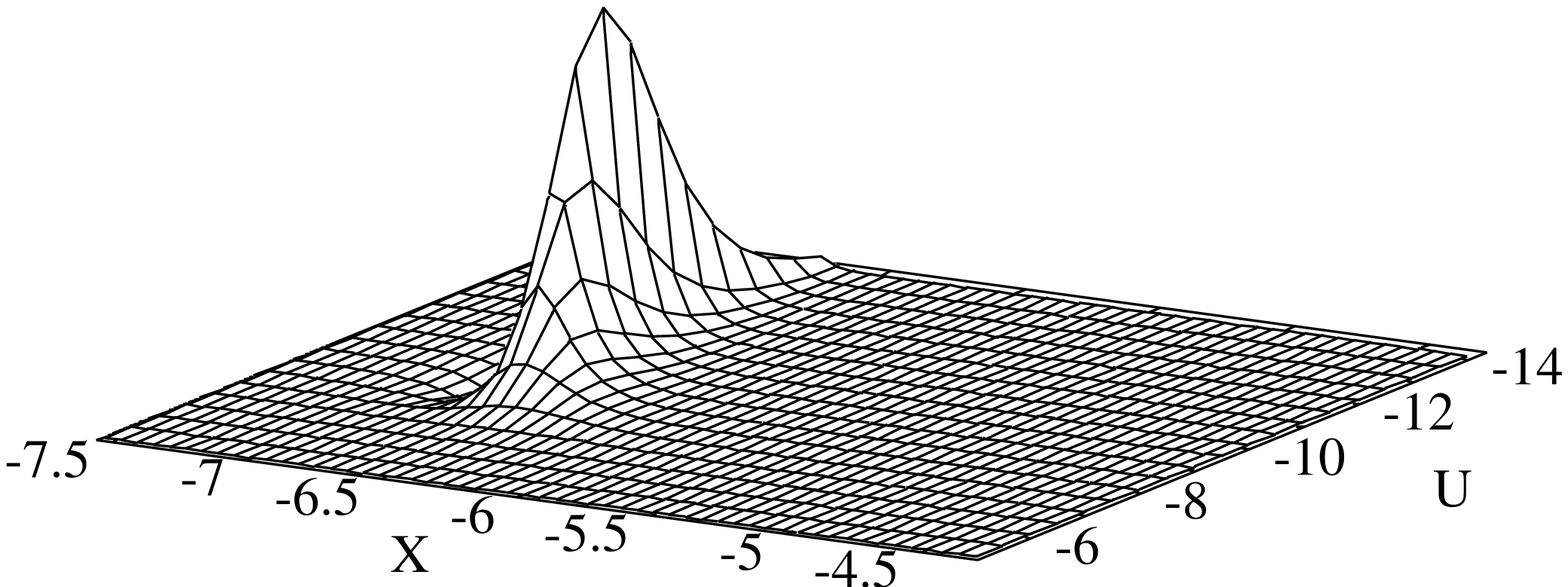}\\
    \ \\
    \hspace*{-1cm} \ (a) \ \hspace{6cm} \ (b) \ \hspace{5cm}
  \end{center}
\caption[]{Integrands of (a) the decay constant
in Eq.~(\ref{eq:F_V}) and (b) the normalization condition
in Eq.~(\ref{eq:normalization})
for $\alpha_\ast=0.885$.
}
\label{fig:sfT0885}
\end{figure}

\begin{figure}
  \begin{center}
    \includegraphics[width=6cm]{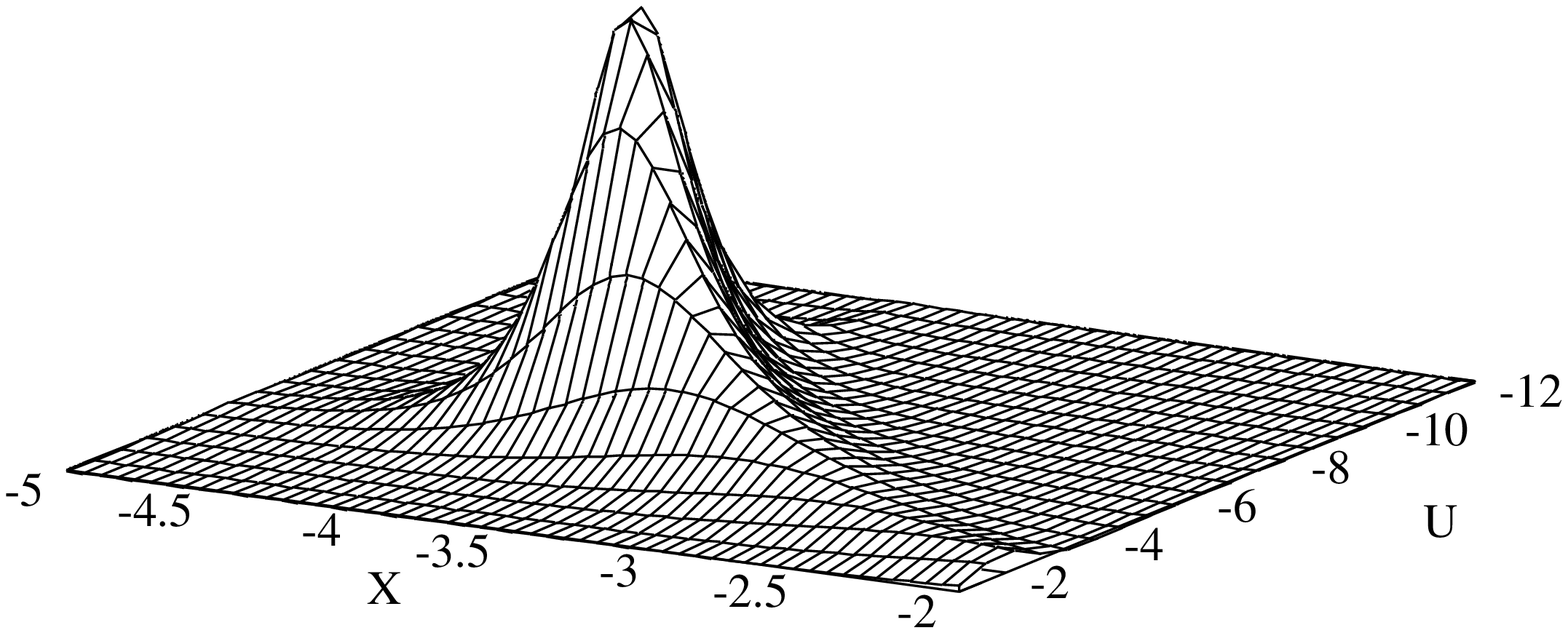}
    \ \includegraphics[width=6cm]{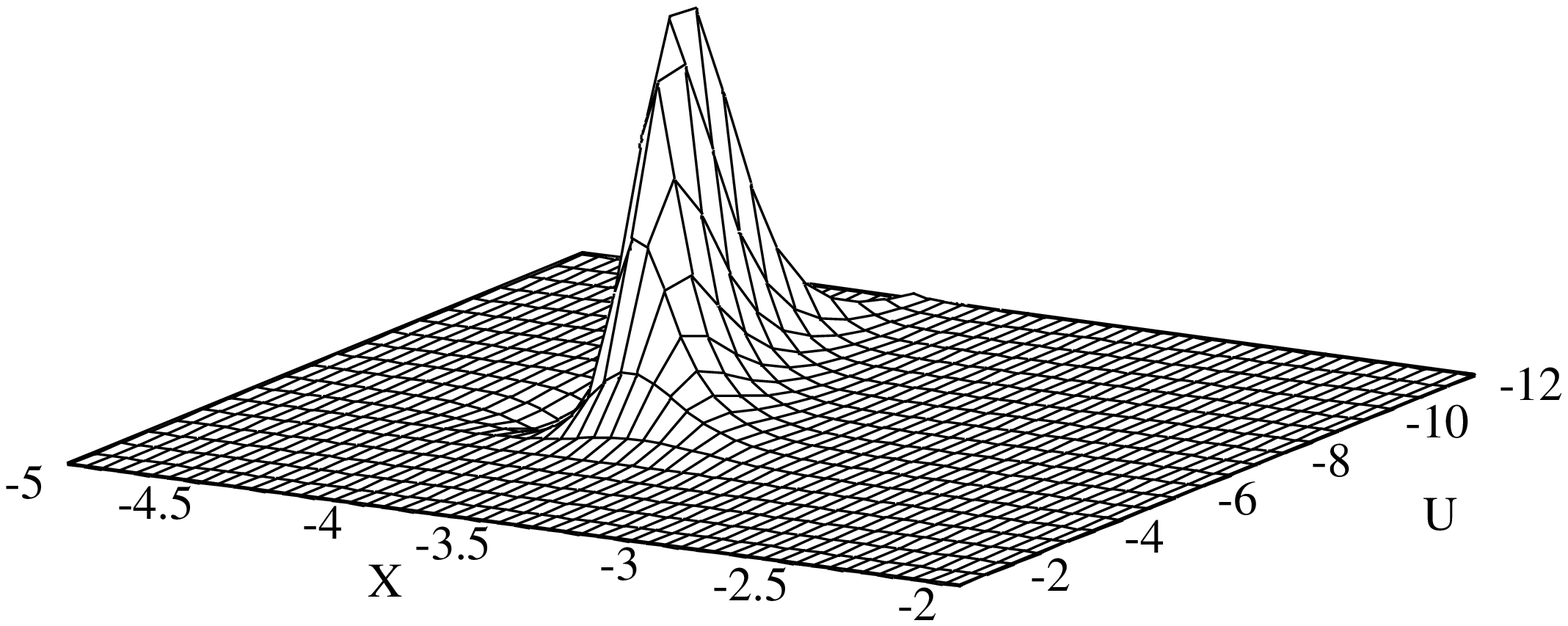}\\
    \ \\
    \hspace*{-1cm} \ (a) \ \hspace{6cm} \ (b) \ \hspace{5cm}
  \end{center}
\caption[]{Integrands of \ (a)\  the decay constant
in Eq.~(\ref{eq:F_V}) and \ (b)\  the normalization condition
in Eq.~(\ref{eq:normalization})
for $\alpha_\ast=1.0$.
}
\label{fig:sfT1000}
\end{figure}

To solve the HBS equation for the bound states numerically,
we introduced the UV and IR cutoffs 
and converted the HBS equation into a linear eigenvalue equation by
discretizing the integral.
As we have discussed in subsection~\ref{ssec:VB},
we adjust the values of the UV and IR cutoffs in such a way that
the dominant supports of the integrands of the decay constant
in Eq.~(\ref{eq:F_V}) and the normalization condition
in Eq.~(\ref{eq:normalization})
lie in the energy region between the UV and IR cutoffs.
In Figs.~\ref{fig:sfT0885} and \ref{fig:sfT1000} we show those
integrands for $\alpha_\ast = 0.885$ and $1.0$ in the case of 
the vector bound state.
These figures show that the dominant supports lie in the lower energy
region for smaller value of $\alpha_\ast$, and that the present choices
in Eqs.~(\ref{eq:IR cutoffs}) and (\ref{eq:UV cutoffs})  
covers the supports.
For other values of $\alpha_\ast$ used in the present analysis
we have checked that the dominant supports always lie within
the energy region between the UV and IR cutoffs chosen as in
Eqs.~(\ref{eq:IR cutoffs}) and (\ref{eq:UV cutoffs}).

\begin{figure}
  \begin{center}
    \includegraphics[height=5cm]{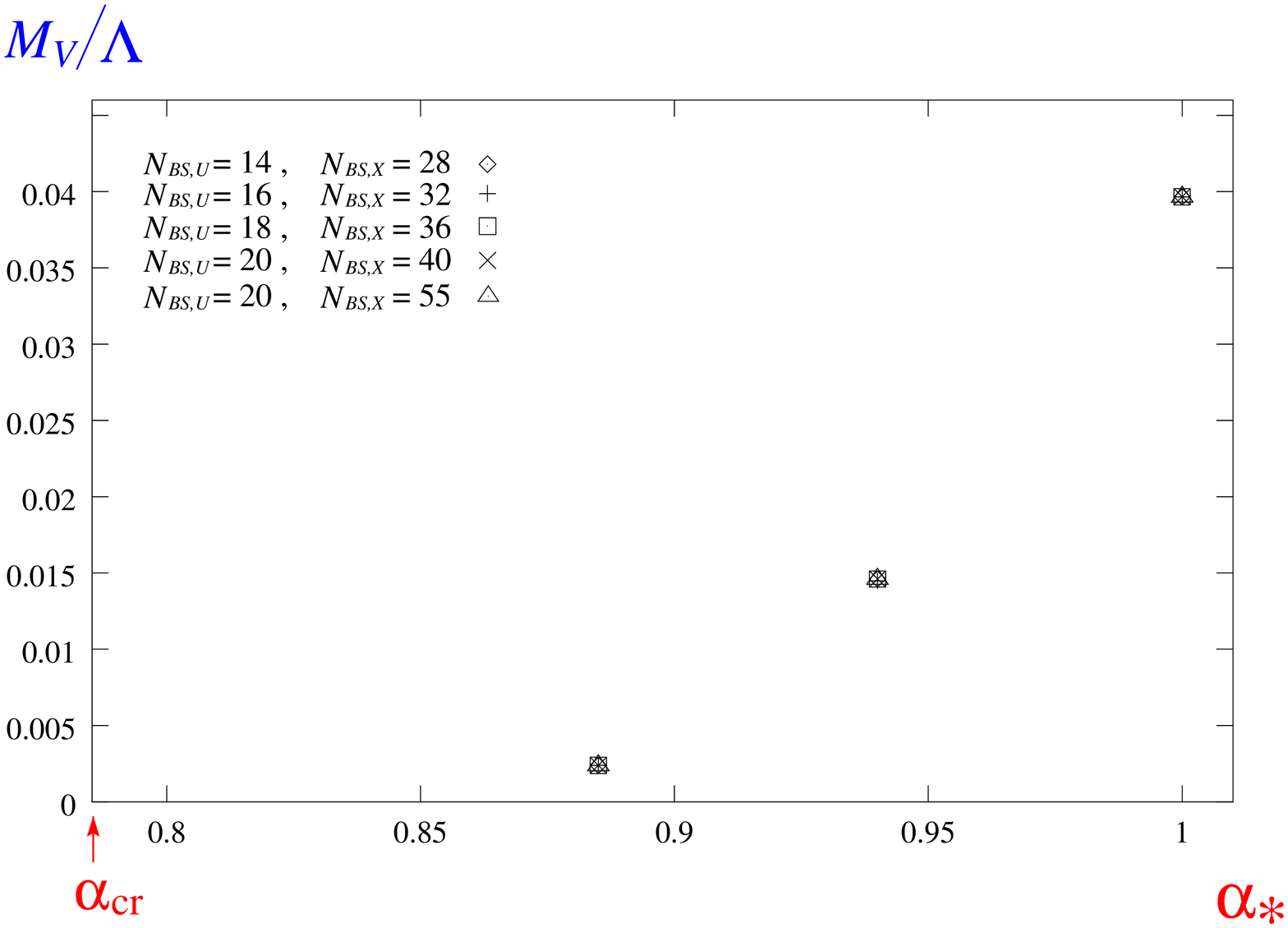} \hspace{1cm}
    \includegraphics[height=5cm]{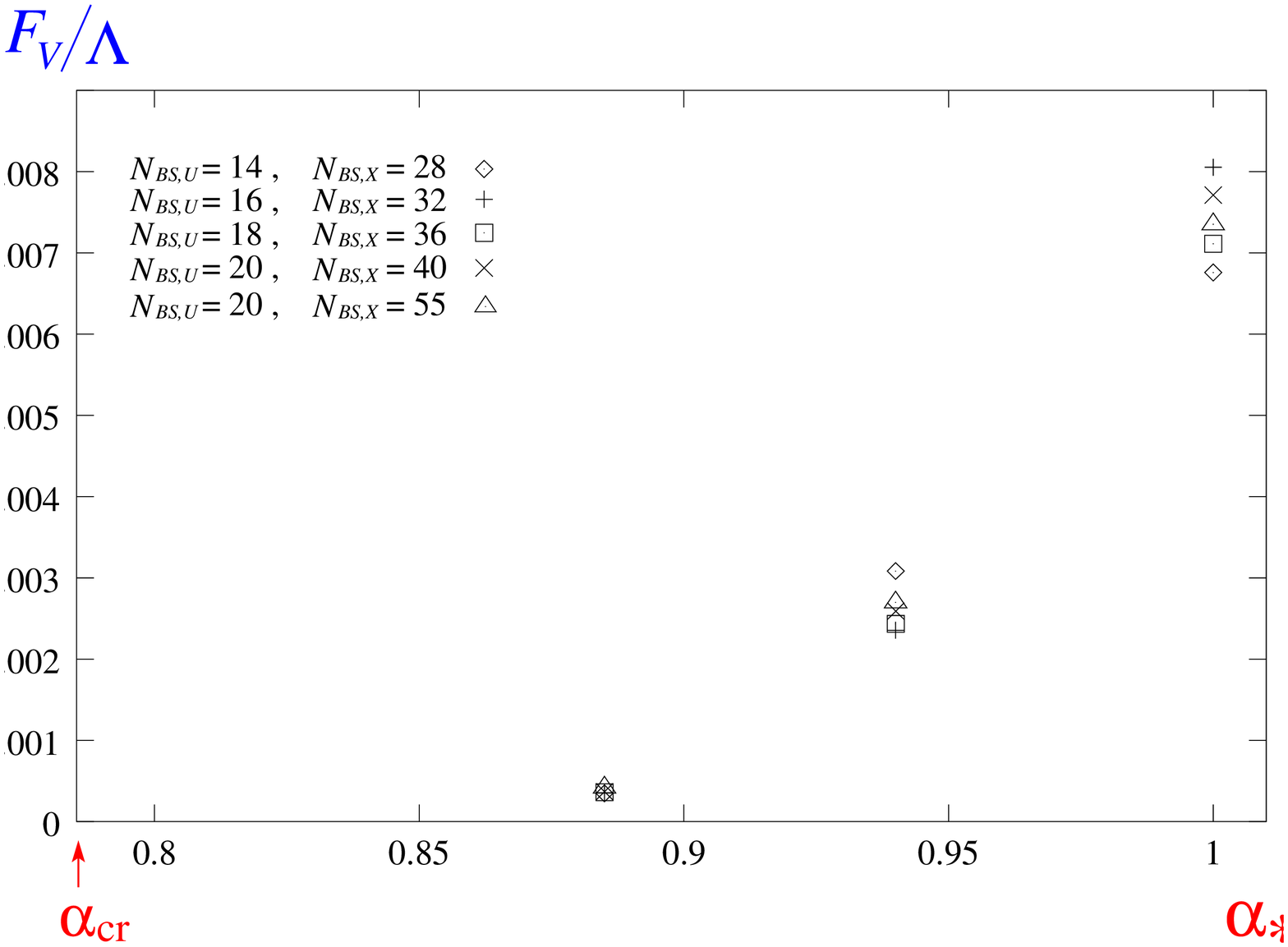}
    \ \\
    \hspace*{0.2cm} \ (a) \ \hspace{7.2cm} \ (b) 
  \end{center}
\caption{Typical values of \ (a)\  $M_V/\Lambda$ and \ (b)\ $F_V/\Lambda$ 
for five choices of  the size of discretization,
$(N_{BS,U},N_{BS,X}) = (14,28)$, $(16,32)$, $(18,36)$, $(20,40)$, 
and $(20,55)$.
}
\label{fig:NBS_dependence}
\end{figure}

As for the numbers of the discretization,
due to the limitation of the computer resources
we used $N_{BS,U} = 20$ and $N_{BS,X} = 55$ as shown in 
Eq.~(\ref{NBS num}).
Here we study the dependences of the mass and the decay constant 
of the vector bound state on the size of discretization.
We show the typical values
of the mass 
in Fig.~\ref{fig:NBS_dependence} (a) and those of
the decay constant in 
Fig.~\ref{fig:NBS_dependence} (b)
for five choices of
the size of the discretization,
$(N_{BS,U},N_{BS,X}) = (14,28)$, $(16,32)$, $(18,36)$, $(20,40)$, 
and $(20,55)$.
Figure~\ref{fig:NBS_dependence} (a) clearly shows that 
the choice $(N_{BS,U},N_{BS,X}) = (20,55)$ is large enough to obtain 
the mass of the vector bound state.
On the other hand,
Fig.~\ref{fig:NBS_dependence} (b) shows that there are still
uncertainties in the value of the decay constant which come from the
size of the discretization.
Apparently, this uncertainty from the size of discretization is the
dominant part of the numerical uncertainties in the present analysis.



\begin{thebibliography}{99}

\bibitem{Banks:nn}
T.~Banks and A.~Zaks,
Nucl.\ Phys.\ B {\bf 196}, 189 (1982).

\bibitem{Appelquist:1996dq}
T.~Appelquist, J.~Terning and L.~C.~Wijewardhana,
Phys.\ Rev.\ Lett.\  {\bf 77}, 1214 (1996)
[arXiv:hep-ph/9602385].

\bibitem{Miransky:1996pd}
V.~A.~Miransky and K.~Yamawaki,
Phys.\ Rev.\ D {\bf 55}, 5051 (1997)
[Erratum-ibid.\ D {\bf 56}, 3768 (1997)]
[arXiv:hep-th/9611142].

\bibitem{Appelquist:1998rb}
T.~Appelquist, A.~Ratnaweera, J.~Terning and L.~C.~Wijewardhana,
Phys.\ Rev.\ D {\bf 58}, 105017 (1998)
[arXiv:hep-ph/9806472].

\bibitem{lattice}
J.~B.~Kogut and D.~K.~Sinclair,
Nucl.\ Phys.\ B {\bf 295}, 465 (1988);
F.~R.~Brown, H.~Chen, N.~H.~Christ, Z.~Dong, R.~D.~Mawhinney,
W.~Schaffer and A.~Vaccarino,
Phys.\ Rev.\ D {\bf 46}, 5655 (1992);
Y.~Iwasaki, K.~Kanaya, S.~Sakai and T.~Yoshie,
Phys.\ Rev.\ Lett.\  {\bf 69}, 21 (1992);
Y.~Iwasaki, K.~Kanaya, S.~Kaya, S.~Sakai and T.~Yoshie,
Nucl.\ Phys.\ Proc.\ Suppl.\  {\bf 53}, 449 (1997);
Prog.\ Theor.\ Phys.\ Suppl.\  {\bf 131}, 415 (1998).

\bibitem{OZ}
R.~Oehme and W.~Zimmermann,
Phys.\ Rev.\ D {\bf 21}, 471 (1980).

\bibitem{VS}
M.~Velkovsky and E.~Shuryak,
Phys.\ Lett.\ B {\bf 437}, 398 (1998).

\bibitem{hot-dense}
T.~Hatsuda and T.~Kunihiro,
Phys.\ Rept.\  {\bf 247}, 221 (1994);
R.~D.~Pisarski,
hep-ph/9503330;
G.E.~Brown and M.~Rho, 
Phys.\ Rept.\  {\bf 269}, 333 (1996);
F.~Wilczek,
hep-ph/0003183;
G.~E.~Brown and M.~Rho,
Phys.\ Rept.\  {\bf 363}, 85 (2002)
[arXiv:hep-ph/0103102].

\bibitem{Harada:1999zj}
M.~Harada and K.~Yamawaki,
Phys.\ Rev.\ Lett.\  {\bf 83}, 3374 (1999)
[arXiv:hep-ph/9906445].

\bibitem{Harada:2000kb}
M.~Harada and K.~Yamawaki,
Phys.\ Rev.\ Lett.\  {\bf 86}, 757 (2001)
[arXiv:hep-ph/0010207].

\bibitem{HY:PRep}
M.~Harada and K.~Yamawaki,
to appear in Phys. Rept.
[arXiv:hep-ph/0302103].

\bibitem{Bando:1984ej}
M.~Bando, T.~Kugo, S.~Uehara, K.~Yamawaki and T.~Yanagida,
Phys.\ Rev.\ Lett.\  {\bf 54}, 1215 (1985);\  
M.~Bando, T.~Kugo and K.~Yamawaki,
Phys.\ Rept.\  {\bf 164}, 217 (1988).


\bibitem{Chivukula:1996kg}
R.~S.~Chivukula,
Phys.\ Rev.\ D {\bf 55}, 5238 (1997)
[arXiv:hep-ph/9612267].

\bibitem{Nakanishi:ph}
N.~Nakanishi,
Prog.\ Theor.\ Phys.\ Suppl.\  {\bf 43}, 1 (1969).

\bibitem{Kugo:review}
T.~Kugo,
in {\it Proc. of 1991 Nagoya Spring School on Dynamical
Symmetry Breaking, Nakatsugawa, Japan, 1991}, ed. K.~Yamawaki  
(World Scientific, Singapore, 1992).

\bibitem{Miransky:book}
V.~A.~Miransky,
``Dynamical symmetry breaking in quantum field theories,''
{\it  Singapore, Singapore: 
World Scientific (1993) 533 p}.

\bibitem{MN}
T.~Maskawa and H.~Nakajima,
Prog.\ Theor.\ Phys.\  {\bf 52}, 1326 (1974);
T.~Maskawa and H.~Nakajima,
Prog.\ Theor.\ Phys.\  {\bf 54}, 860 (1975);
M.~Bando, M.~Harada and T.~Kugo,
Prog.\ Theor.\ Phys.\  {\bf 91}, 927 (1994)
[arXiv:hep-ph/9312343].

\bibitem{Kugo:1992pr}
T.~Kugo and M.~G.~Mitchard,
Phys.\ Lett.\ B {\bf 282}, 162 (1992).

\bibitem{Kugo:1992zg}
T.~Kugo and M.~G.~Mitchard,
Phys.\ Lett.\ B {\bf 286}, 355 (1992).

\bibitem{Bando-Harada-Kugo}
M.~Bando, M.~Harada and T.~Kugo,
Prog.\ Theor.\ Phys.\  {\bf 91}, 927 (1994)
[arXiv:hep-ph/9312343].

\bibitem{Higashijima:1983gx}
K.~Higashijima,
Phys.\ Rev.\ D {\bf 29}, 1228 (1984).

\bibitem{Miransky:vj}
V.~A.~Miransky,
Sov.\ J.\ Nucl.\ Phys.\  {\bf 38}, 280 (1983)
[Yad.\ Fiz.\  {\bf 38}, 468 (1983)].

\bibitem{Aoki:1990eq}
K.~I.~Aoki, M.~Bando, T.~Kugo, M.~G.~Mitchard and H.~Nakatani,
Prog.\ Theor.\ Phys.\  {\bf 84}, 683 (1990).

\bibitem{Aoki:1990aq}
K.~I.~Aoki, M.~Bando, T.~Kugo and M.~G.~Mitchard,
Prog.\ Theor.\ Phys.\  {\bf 85}, 355 (1991).


\bibitem{Roberts-Schmidt}
C.~D.~Roberts and S.~M.~Schmidt,
Prog.\ Part.\ Nucl.\ Phys.\  {\bf 45}, S1 (2000)
[arXiv:nucl-th/0005064].

\bibitem{Kugo-Mitchard-Yoshida}
T.~Kugo, M.~G.~Mitchard and Y.~Yoshida,
Prog.\ Theor.\ Phys.\  {\bf 91}, 521 (1994)
[arXiv:hep-ph/9312267].

\bibitem{Maris-Roberts-Tandy}
P.~Maris, C.~D.~Roberts and P.~C.~Tandy,
Phys.\ Lett.\ B {\bf 420}, 267 (1998)
[arXiv:nucl-th/9707003].

\bibitem{Aoki:1990yp}
K.~I.~Aoki, T.~Kugo and M.~G.~Mitchard,
Phys.\ Lett.\ B {\bf 266}, 467 (1991).

\bibitem{Naito-Oka}
K.~Naito, K.~Yoshida, Y.~Nemoto, M.~Oka and M.~Takizawa,
Phys.\ Rev.\ C {\bf 59}, 1095 (1999)
[arXiv:hep-ph/9805243];
K.~Naito and M.~Oka,
Phys.\ Rev.\ C {\bf 59}, 542 (1999)
[arXiv:hep-ph/9805258].

\bibitem{Alkofer-Watson-Weigel}
R.~Alkofer, P.~Watson and H.~Weigel,
Phys.\ Rev.\ D {\bf 65}, 094026 (2002)
[arXiv:hep-ph/0202053].

\bibitem{Gardi}
E.~Gardi and M.~Karliner,
Nucl.\ Phys.\ B {\bf 529}, 383 (1998)
[arXiv:hep-ph/9802218];
E.~Gardi, G.~Grunberg and M.~Karliner,
JHEP {\bf 9807}, 007 (1998)
[arXiv:hep-ph/9806462].

\bibitem{Leung-Love-Bardeen}
C.~N.~Leung, S.~T.~Love and W.~A.~Bardeen,
Nucl.\ Phys.\ B {\bf 273}, 649 (1986);
Nucl.\ Phys.\ B {\bf 323}, 493 (1989).


\bibitem{Appelquist:1988yc}
T.~Appelquist, K.~D.~Lane and U.~Mahanta,
Phys.\ Rev.\ Lett.\  {\bf 61}, 1553 (1988);
H.~Georgi, E.~H.~Simmons and A.~G.~Cohen,
Phys.\ Lett.\ B {\bf 236}, 183 (1990).

\bibitem{ExplicitSolution}
R.~M.~Corless, G.~H.~Gonnet, D.~G.~E.~Hare, D.~J.~Jeffrey and D.~E.~Knuth,
Adv.\ Comput.\ Math.\  {\bf 5}, 329 (1996).

\bibitem{Pagels:hd}
H.~Pagels and S.~Stokar,
Phys.\ Rev.\ D {\bf 20}, 2947 (1979).

\bibitem{Yamawaki:1985zg}
B. Holdom,
Phys.\ Lett.\ B {\bf 150}, 301 (1985);
K. Yamawaki, M. Bando and K. Matumoto,
Phys.\ Rev.\ Lett.\  {\bf 56}, 1335 (1986);
T. Akiba and T. Yanagida,
Phys.\ Lett.\ B {\bf 169}, 432 (1986);
T.W. Appelquist, D. Karabali and L.C.R. Wijewardhana,
Phys.\ Rev.\ Lett.\  {\bf 57}, 957 (1986).

\bibitem{Kugo-Yoshida}
T.~Kugo and Y.~Yoshida,
Soryushiron Kenkyu {\bf 91}, B26 (1995); 
Y.~Yoshida,
Ph.D thesis, Kyoto University (1995) 

\bibitem{Harada:1995nx}
M.~Harada and Y.~Yoshida,
Phys.\ Rev.\ D {\bf 53}, 1482 (1996)
[arXiv:hep-ph/9505206].

\bibitem{Gilman:1967qs}
F.~J.~Gilman and H.~Harari,
Phys.\ Rev.\  {\bf 165}, 1803 (1968).

\bibitem{Weinberg:hw}
S.~Weinberg,
Phys.\ Rev.\  {\bf 177}, 2604 (1969).

\bibitem{Weinberg:SR}
S.~Weinberg,
Phys.\ Rev.\ Lett.\  {\bf 18}, 507 (1967).

\bibitem{KSY91}
K.-I.~Kondo, S.~Shuto and K.~Yamawaki,
Mod.\ Phys.\ Lett.\ A {\bf 6}, 3385 (1991).
  
\bibitem{Aoki}
K.~I.~Aoki, K.~Morikawa, J.~I.~Sumi, H.~Terao and M.~Tomoyose,
Prog.\ Theor.\ Phys.\  {\bf 102}, 1151 (1999)
[arXiv:hep-th/9908042].

\bibitem{Itzykson}
See, for example, C.~Itzykson and J.~B.~Zuber,
{\it Quantum Field Theory} 
(McGraw-Hill, New York, 1980).

\end{thebibliography}
\end{document}